\documentclass[3p]{elsarticle}

\journal{Pervasive and Mobile Computing}
\date{November 5, 2016}

\usepackage{amsmath}
\usepackage{amssymb}
\usepackage{amsfonts}
\usepackage{amsthm}
\usepackage{mathtools}
\DeclareMathOperator*{\argmin}{argmin}
\usepackage{algorithm}
\usepackage{algorithmic}
\usepackage{array}

\usepackage[caption=false,font=footnotesize]{subfig}

\newcommand{\bs}[1]{\boldsymbol{#1}}

\usepackage{graphicx}
\graphicspath{{./img/}}
 \DeclareGraphicsExtensions{.pdf,.jpeg,.png}

\usepackage[]{changes}
\definechangesauthor[color=red]{AP}

\usepackage{soul}

\sethlcolor{yellow}

\usepackage[]{todonotes}
\newcommand{\cmmnt}[1]{\ignorespaces}

\begin{document}

\begin{frontmatter}

\title{A communication efficient distributed learning framework for smart environments}

\author[iit]{Lorenzo Valerio\corref{cor}}
\cortext[cor]{Corresponding authors}
\ead{l.valerio@iit.cnr.it}

\author[iit]{Andrea Passarella}
\ead{a.passarella@iit.cnr.it}

\author[iit]{Marco Conti}
\ead{m.conti@iit.cnr.it}

\address[iit]{Insitute of Informatics and Telematics, National Research Council, Via Moruzzi 1, Pisa. Italy}

\begin{abstract}

Due to the pervasive diffusion of personal mobile and IoT devices, many ``smart environments'' (e.g., smart cities and smart factories) will be, among others, generators of huge amounts of data. To provide value-add services in these environments, data will have to be analysed to extract knowledge.
Currently, this is typically achieved through centralised cloud-based data analytics services.  However, according to many studies, this  approach may present significant issues from the standpoint of data ownership, and even wireless network capacity.   
One possibility to cope with these shortcomings is to move data analytics closer to where data is generated.  
In this paper we tackle this issue by proposing and analysing a distributed learning framework, whereby data analytics are performed at the \emph{edge} of the network, i.e., on locations very close to where data is generated. Specifically, in our framework, \emph{partial} data analytics are performed directly on the nodes that generate the data, or on nodes close by (e.g., some of the data generators can take this role on behalf of subsets of other nodes nearby). Then, nodes exchange partial models and refine them accordingly. Our framework is general enough to host different analytics services. In the specific case analysed in the paper we focus on a learning task, considering two distributed learning algorithms.
Using an activity recognition and a pattern recognition task, both on reference datasets, we  compare the two learning algorithms between each other and with a central cloud solution (i.e., one that has access to the complete datasets). Our results show that using distributed machine learning techniques, it is possible to drastically reduce the network overhead, while obtaining performance comparable to the cloud solution in terms of learning accuracy. The analysis also shows when each distributed learning approach is preferable, based on the specific distribution of the data on the nodes. 

\end{abstract}

\begin{keyword}
Iot, big data, smart cities, distributed learning, communications efficiency 
\end{keyword}

\end{frontmatter}


\section{Introduction} \label{sec:intro}

\subsection{Motivation}\label{sub:motivation}
The massive diffusion of personal mobile and IoT devices is one of the propellers behind many ``smart environments'' where pervasive networking and computing technologies are increasingly applied. The sheer number of devices predicted to be installed in smart environments is impressive, commonly considered in the order of billions or trillions by 2025, resulting in a density of about 1000 devices per person all over the world~\cite{Vincentelli:2015aa}. Perhaps even more importantly, each of these devices is a data generator, and their pervasive diffusion means that huge amounts of data will be generated at the \emph{edge} of the network, where these devices will reside. According to Cisco, also as a side effect of the pervasive diffusion of personal mobile and IoT devices, the global mobile data traffic will increase at an exponential pace. It will increase nearly eightfold between 2015 and 2020, at a compound annual growth rate (CAGR) of 53\%, reaching 30.6 exabytes per month by 2020~\cite{Cisco:sp}. This data traffic increase is typically referred to as the \emph{data tsunami}. These data
 once properly processed, open uncountable possibilities in terms of novel applications and services.
Data and knowledge extracted from data is indeed the ``glue'' bonding together the
physical part of a smart environment (e.g., the physical infrastructures of a city, such as lighting,
heating, traffic control, water distribution, smart grids, and, extensively,
people) and its cyber part (i.e., the Internet services based on its data). The continuous
interaction between the cyber and the physical parts, mediated by data, enables
not only to sense the status of the physical environment, but also to act upon it based on the
knowledge extracted from data. This is at the basis of the Cyber-Physical
convergence concept~\cite{CPW} for pervasive environments.

According to current IoT/M2M paradigms~\cite{Borgia20141,Cavalcante:2016aa,Borgia:2016aa},  the typical
workflow adopted to process and extract knowledge from 
data is based on central cloud systems. Precisely, to extract knowledge from raw data, typical IoT/M2M
architectures (e.g., the ETSI M2M architecture~\cite{ETSI-arch}) consider three
main layers.  In the first layer gateways collect data from individual
``things''. In the second layer, data is sent to the central cloud via
some pervasive (wireless) network infrastructure, such as LTE. In the third layer, knowledge is extracted from data
and provided to applications or
further exploited to create new services.  

Although such IoT/M2M architecture has proved to be very powerful
in current applications,
 it may have
significant shortcomings in the medium term. First, it assumes the presence of a very broadband wireless access
network to move data from the collection to the processing layers. 
Unfortunately, even the adoption of the latest 4G LTE-A technologies might not
provide sufficient capacity in the medium term~\cite{Rebecchi:2014ys,Valerio:2015ab,Rebecchi:2015aa}. 
Right now we are in the early adoption phase of 4G
technologies, which is giving a significant boost to the cellular capacity with
respect to the current demands.  However, projections over the next  years
show that this additional capacity may be saturated soon and may not be sufficient to cope with
the data tsunami effect~\cite{Cisco:sp}. As a consequence, bandwidth crunch
events similar to those occurred with 3G networks~\cite{NYTimes:fp} may be
expected, unless alternative solutions are devised in time. Second,
  collecting all data in a central place might not be always feasible due to
  privacy or ownership constraints~\cite{Estrada-Jimanez:2017aa}. This applies both to general personal
  devices of users, who might not be willing to give away data about their
  personal behaviour, but also to many ``verticals'' such as industrial
  environments, where significant fractions of the data would have to remain
  local to specific production environments~\cite{5GPPP-FoF}.

One possibility to cope with these problems is to limit
the amount of data sent to global cloud platforms, and move part or the data
processing tasks to the edge of the network, where data is generated.  This is
the vision of (Mobile) Edge Computing~\cite{Lopez:2015aa},  Fog
computing \cite{Bonomi:2012aa}, and
opportunistic computing \cite{Conti:2010:ONO:1866991.1867009}. Not only
this approach can help offloading excessive
traffic from wireless access networks.  It can indeed be the most suitable
approach to extract knowledge also from privacy sensitive data, which may not be
transferred to third parties (global cloud operators) for
processing~\cite{Lopez:2015aa}. Note that this approach is gaining more and more momentum both in the research and industry environments. For example, Google has recently advertised a mobile computing framework, called Federated Learning~\cite{McMahan:2017aa}. In Federated Learning, mobile devices (smartphones or tablet) spread all over the world, collaborate to perform a deep learning task and  learn a global  model from their local data. All the learning phases are incrementally accomplished directly on the data owned locally by devices, and data is never sent to the cloud.

This approach can be applied to multiple instances of smart environments. One example are smart factories, where data about the production process, collected either by sensors or by operators devices, need to be analysed to understand the status of the production, in order to dynamically tune or even reconfigure the processes. This is the typical case considered in the Industry 4.0 (or Industrial Internet) framework~\cite{Kagermann:2013aa}. In most of the cases, industries are not willing to move their data to some external cloud provider infrastructure, due to confidentiality reasons. On the other hand, they will not have the competences to build and manage a private cloud platform. In this case, decentralised analytics solutions where data are directly analysed on the devices available on the factory premises will have several advantages. It will avoid leaking of information outside of the factory environment, and will reduce the dependency and cost of IT services provided by external IT operators~\cite{5GPPP-FoF}.
Another use case for distributed analytics are more general smart cities services. For instance, think to a mobile
crowd sensing scenario in which people located in a limited geographical area
(e.g. a park, square, museum, etc.) are performing some activity, and their
smart devices have to infer collectively what, either single users or the
  whole crowd, is doing. Privacy issues may restrict the
  transmission of data to a central cloud \cite{Teles:2017aa}. Moreover, most of
the time analytics computed over these data may be relevant only for limited time
span and at very specific locations, making movement of data back and
forth the devices in those locations a complete waste of bandwidth. This
  is the scenario for which several localised mobile social networking
  applications have recently been created (e.g.,
  Here\&Now~\cite{Karkkainen:2016aa}, Crowd-Cache~\cite{Thilakarathna:2017aa}). In this context, additional trust and incentives issues may arise, which are typically not present in smart factory environments (or other environments where the set of devices is under the ownership of a single or a federation of operators). For example, users would need to be incentivised to perform data analytics on behalf of each other, and they need to trust the results computed by the others.

\subsection{Contribution}\label{sub:contribution}
In this paper we focus on the mobile crowd sensing
scenario, and we target the general problem of extracting knowledge through
machine learning techniques from a dataset that is collected across a number of
physical locations, and available at distinct mobile nodes, one per
location\footnote{Indeed, in the following we will use the concept of physical
    location and mobile node storing data collected at that location
interchangeably.}. 
 We define a framework where  learning is mostly performed locally at each
 individual location, and the  amount of information to be shared across
 locations is minimised. Precisely, in our case we compute partial models built on
 the local datasets and our purpose is to come up with a more refined
 model that includes the knowledge of all the partial models, while
   minimising the amount of information exchanged between nodes that compute the
 partial models. 
 
We focus on two real learning cases, and use reference datasets in
   the literature. The first one is about activity recognition from data sampled by smart-phones in
   individual locations. The second one is about digits recognition.
   Note that the main purpose of the datasets is to show, with reference
   benchmarks the validity of the proposed distributed learning approach, more than providing
   specific use cases for smart cities scenarios. As a matter of fact, the first
   dataset serves also the latter role. It is, in fact, representative of cases
   where users' personal devices need to recognise activities that are
   contextualised to the specific behaviour of their users in specific
   locations, possibly for limited amounts of time. In these cases, recognising
   the activity of each individual user by monitoring exclusively its own
   behaviour (a typical activity recognition task which can implemented at each
   node in isolation) may not be meaningful, as the activities would be specific
   to the location and context of the user during a certain time window.
   It is more useful to derive activity recognition models from
   a multitude of users who would behave approximately the same because they
 share the same context for a certain time. There are a few important remarks bound to this scenario. The first one is related to data labelling. We point out that in this paper we focus on the definition and evaluation of the leaning framework, assuming that data held by phones has been already labelled either by the user or other  mechanisms. Manual tagging for collected data such as in the  activity recognition example would not be totally uncommon in these contexts. A second remark is related to incentives for users to collaborate in the distributed task, and trust mechanisms by which users can trust the partial models computed by others. Both problems are very relevant, but they are common to several crowd sensing systems. In this paper we consider them mostly as orthogonal problems. Note, however, that in the paper we analyse the robustness of our scheme to malicious behaviours, and we show that the proposed scheme is able to automatically filter out maliciously modified partial models. This property, together with dedicated mechanisms, can help in addressing trust issues.

As in our prior work~\cite{Valerio:2016aa}, in this paper we use a distributed learning algorithm based on the Hypothesis Transfer Learning (HTL) scheme proposed in~\cite{OrabonaGreedyTL}. The algorithm we use trains a separate model for each location where part of the dataset is available, exchanges partial models between locations, and finally obtains a unique refined model.
 In this paper, we significantly extend our prior work in several directions. We consider a
   set of different distributed learning models, in addition to HTL. We contrast
   the prediction performance of all the schemes between each other, and with
   respect to a centralised solution where all data is sent to a central cloud
   platform. Specifically, we compare all schemes in terms of prediction
   performance and network overhead.
   Our results show that, in general all distributed solutions achieve
   prediction accuracies very close to the centralised cloud solution.
   Depending on the scenario and
   the distribution of data on the nodes, different distributed learning
   approaches yield the best trade-off between accuracy and network traffic.  However, in
   all cases, the best distributed approach yields accuracy comparable with a
   centralised solution, while drastically cutting the network traffic, between
   52\% and up to 99\%.   
   We also derive analytical bounds on this gain, and analyse its
   sensitiveness with respect to the key involved parameters. Moreover, we
   compare the distributed solutions in terms of the trade off between network
   traffic and prediction performance. In general, HTL outperforms simpler
   solutions in terms of prediction by generating higher network traffic.
   Therefore, we define a modified HTL solution where traffic is drastically
   reduced, without compromising the prediction performance. In addition, we evaluate the performance of all the distributed solutions in presence of malicious users, which inject wrong partial models in the system. We show that, contrarily to other distributed algorithms, HTL is very robust in this case, as nodes are able to automatically filter out corrupted partial models. Finally, we analyse the performance of the distributed algorithms in dynamic scenarios, where nodes come and go dynamically over time, and participate to the learning task during different time intervals. We show that also in this case distributed algorithms can achieve very high accuracy with a drastic reduction of network traffic.
   All in all, we think that these results provide a strong validation on the viability of distributed data analytics schemes at the edge of the network. In the paper we consider learning problems, i.e., a very specific instance of data analytics, to have a practical case to work with. However, the framework we propose is more general, and can accommodate any data analytics problem for which distributed algorithms are available.

 The rest of the paper is organised as follows. Section~\ref{sec:related}
 presents related work.
 Section~\ref{sec:TL} provides a brief background on
 HTL, which is then applied to our problem in
 Section~\ref{sec:distributed-learning}. Sections~\ref{sec:metrics}
 describe the datasets we have used. Section~\ref{sec:peva} compares the
 distributed solutions with each other and with the cloud-based solution.
 Section~\ref{sec:robust} shows the robustness of HTL to malicious behaviour of devices. 
 Section~\ref{sec:noh} analyses the network traffic overhead, while
 Section~\ref{sec:sensitivity} shows how to further optimise the trade-off between
 traffic overhead and prediction accuracy for HTL. 
 Section~\ref{sec:dyn} analyses the performance of the distributed learning algorithms in dynamic conditions.
 Finally,
 Section~\ref{sec:conclusions} concludes the paper.

\section{Related work}
\label{sec:related}
The distributed learning problem could be approached in many ways. One
possibility is represented by ``ensemble methods'' such as bagging, boosting and
mixtures of experts~\cite{Breiman:aa, Freund:1997aa, Freund:1997ab,
Jacobs:1991aa}. Intuitively, these methods allocate portions of the training
dataset to different classifiers that are independently trained. The individual
classifiers are subsequently aggregated, sometimes through a different training
algorithm and sometimes using feedback from the training phase of the individual
classifiers. These approaches might be suitable for distributed learning in
parallel networks, however they are generally designed within the classical
model for supervised learning and fundamentally assume that the training set is
available to a single coordinating processor.

Motivated by the presence of huge quantities of data, there are other machine
learning techniques that focus on scaling up the typical centralized learning
algorithm. One approach is to decompose the training set into smaller ``chunks''
and subsequently parallelize the learning process by assigning distinct
processors/agents to each of the chunks. This is the typical scenario in which
deep learning is widely used~\cite{Bordes:2005aa, Dean:2012aa, Coates:2013aa}.
Differently from our solution, these approaches do not target knowledge
extraction where data have privacy constraints, or when network overhead should
be minimised.

Finally, other works propose  fully distributed and decentralized
learning algorithms. To the best of our knowledge, the  more relevant with
respect to our reference scenario are the following.  In~\cite{navia2006distributed}
authors present a distributed version of Support
Vector Machine (SVM). In order to learn a model from physically distributed
data, they propose an iterative procedure according to which they i) train an
SVM on each local dataset and then they exchange the learned Support Vectors (i.e.
relevant points of the dataset that determine the classifier) with all the other
locations. These Support Vectors are added to each local dataset and the local
SVMs are trained again.  Authors of~\cite{Georgopoulos20142} propose a
distributed learning algorithm for arbitrary connected machines. The approach
consists of an initial learning phase performed on the local data and then an
iterative consensus procedure through which machines come up with a final model.
Another similar solution presented in~\cite{Scardapane2015271} propose two
distributed learning algorithms for Random Vector Functional Link Networks (RVFL). The first
one is based on a Decentralised Consensus Algorithm. The second one, instead, is
based on the Alternative Direction Method of Multipliers. Both are iterative
solutions that in order to converge to a model have to repeatedly exchange
their partial models' parameters. 
Other solutions belong the ``stacked generalisation'' family whose the typical approach is to learn a global model that combines the outputs of several partial models~\cite{Wolpert:1992aa}. Precisely, in this approach every machine holds a local dataset that is split into training set and validation set. Each machine trains a local model on training set and share it with all the other locations.
At each location all the  received models are tested on the validation set and answers are sent to a unique collecting machine that trains a ``meta-model'' on this newly created dataset. This last trained model represents the final model. 
This approach is not suitable in our case because it would generate an amount of traffic that is proportional to the size of the validation sets at each location and to the number of locations.

All these solutions typically converge to a model whose accuracy is
comparable with a centralized algorithm with access to the entire dataset.
However, none of them consider the amount of network traffic triggered by
their iterative procedures. 
In this paper, instead, we consider two state of the art ensemble learning approaches that are suitable to be trained without a complete ``view'' of the dataset and we compare their performance with a distributed learning solution based on Hypothesis Transfer Learning proposed  in our  previous paper~\cite{Valerio:2014qf}. It is worth noting that although it shares some similarity with the  ``stacking generalisation'' paradigm, it does not send any data over the network. Moreover, for the sake of completeness we compare the performance of all the considered procedures with a centralised cloud-based benchmark, i.e. a learning algorithm that has complete access to the entire dataset at hand.

The main novelty and goal of this paper is not to champion one distributed learning approach over the others, but rather to perform a comparative study that highlights, in different working scenarios, their benefits. Our purpose, in the end, is to be able to select the best learning technique, in terms of accuracy and networks traffic, for different IoT scenario.

\section{Hypothesis transfer learning}
\label{sec:TL}

In this section we present the core part of the distributed learning solution based on Hypothesis Transfer Learning. Specifically,  in this paper the general learning task we take into account is about classification. Therefore, the Hypothesis Transfer Learning approach we selected to build the distributed learning approach is a classification algorithm called GreedyTL~\cite{OrabonaGreedyTL}. 

For the sake of clarity, before providing the details of GreedyTL, we describe with a simple example a high level intuition about the general Hypothesis Transfer Learning framework.
Let us consider the existence of one small dataset $D_1$ and two larger datasets $D_2,D_3$ collected across three separate locations and let us consider that these datasets are only locally accessible. 
Let us suppose that we already have learnt a model  for both $D_2$ and $D_3$, here denoted by $h_2$ and $h_3$, respectively.
Now we want to learn a model on $D_1$, denoted by $h_1$. Note that, since in machine learning the accuracy of a model strongly depends on the size of the dataset, the small size of $D_1$ will negatively affect the accuracy of $h_1$. According to the HTL framework, in order to overcome this problem we can exploit the knowledge contained in $h_2,h_3$ to learn a more accurate model $h_1$  on the dataset $D_1$. In the HTL terminology, the models $h_2,h_3$ are called \emph{source models} while $h_1$ is called \emph{target model}. Analogously, $D_2,D_3$ are referred as source domains and $D_1$ as target domain. 

Let us now introduce the notation used in this section: small bold letters denote column vectors. The training set of cardinality $m$ is denoted as the set of pairs $\{(\bs{x}_i,y_i) \}_{i=1}^m$ where $\bs{x}_i \subseteq \mathcal{X} \in \mathbb{R}^d$ and $y \subseteq \mathcal{Y} \in \mathbb{R}$. Here $\mathcal{X}, \mathcal{Y}$ denote respectively the input and the output space of a learning problem. Here we denote with $\{h^{src}_i(\bs{x})\}_{i=1}^L$ the set of $L$ source models and with $h^{trg}$ the target model that we have to learn in the target domain.

GreedyTL focuses on the binary classification problem. However, it is worth noting that this is not a limitation because it can be easily extended to the multi-class classification problem (details about how we used it in a multi-class classification problem are provided in the next section). The model that GreedyTL learns is defined as follows:

\begin{equation}
    h^{trg}(\bs{x}) = \bs{\omega}^T \bs{x} + \sum_{i=1}^L \beta_i h^{src}_i(\bs x)
    \label{eq:htrg}
\end{equation}
Note that the target model is a standard linear classifier of the form $$h^{trg}(\bs x)= \bs\omega^T \bs x$$ with an additional term  $$\sum_{i=1}^L \beta_i h^{src}_i(\bs x)$$ that permits to include the knowledge of the source models into the target model. The ${\bs \beta}$  coefficients control how much each source model can affect the behaviour of the target model. Here  ${\bs\omega}$ and ${\bs \beta}$ are the parameters of the target model to be learnt.

In order to find the best combination of parameter $\bs{\omega}$ and
$\bs{\beta}$, GreedyTL solves the following optimization problem

\begin{eqnarray}
    \label{eq:optprob}
    (\bs{\omega}^*,\bs{\beta}^*) &=  \rm{argmin}_{\bs{\omega,\beta}} \left\{ \hat{R}(h^{trg}) + \lambda \|\bs{\omega}\|^2_2 + \lambda\|\bs{\beta}\|^2_2 \right\} \nonumber\\
    & \bs{s.t.} \|\bs{\omega}\|_0 + \|\bs{\beta}\|_0 \leq \kappa 
\end{eqnarray}
The meaning of the objective function is the following. The first  term $$\hat{R}(h^{trg}) = \frac{1}{m}\sum_{i=1}^m (h^{trg}(\bs{x_i}) -y_i)^2$$ is the mean squared error and says that we are looking for linear classifier $h^{trg}$ that minimises the number of prediction errors. The terms $$\lambda \|\bs{\omega}\|^2_2 + \lambda\|\bs{\beta}\|^2_2$$ are called regularization terms, and their effect is to avoid to find a model that over-fits the data in the training set. This is called Tikhonov regularization an it is a common practice in the machine learning field to improve the generalisation capabilities of the learning algorithms~\cite{Ng:2004aa}. Here $\lambda$ is a parameter used to control the impact of this second term in the final solution of the optimization problem.
Finally, the constraint of the optimization problem $$\|\bs{\omega}\|_0 + \|\bs{\beta}\|_0 \leq \kappa $$imposes that the number of non-null coefficients of the vectors ${\bs\omega}$ and ${\bs \beta}$, collectively do not exceed a constant parameter $\kappa$. In this way we force the algorithm to select only those source models that collectively improve the generalisation of the target model and, as a consequence, the size of the target model is limited.  Summarising in this way we are looking for model of size at most $\kappa$, that minimises the prediction error. 

It is important to note that the formulation of this optimisation problem is a special case of the Subset Selection problem~\cite{das2008algorithms} which aims at finding the subset of $\kappa$ out of $d+L$\footnote{$d+L$ is the sum of the sizes of the $\omega$ and $\beta$ vectors, respectively.} variables that maximises the prediction of a selected variable (such as the class label for a classification problem)\cite{OrabonaGreedyTL}. However, as specified in~\cite{das2008algorithms} the subset selection problem is NP-hard, therefore we can only find an approximate solution. To this end the problem (\ref{eq:optprob}) is solved using a regularized Least Square Forward Regression algorithm that iteratively tests and adds features to the solution until the number of features added is  $\leq \kappa$ and such that the prediction error is minimised. 

As shown in \cite{OrabonaGreedyTL}, this algorithm is very accurate even with very small-sized training sets. From a computational point of view, the small size of the training set, however is a mandatory condition for GreedyTL because it has to invert a matrix whose size is proportional to the size of the local dataset at hand, i.e. the matrix inversion operation is computationally very expensive. 
Since we want our procedure to be general, we cannot pose a limit on the size of the dataset. Therefore, we overcome this limitation in the following way. For each data location we train several instances of GreedyTL on different randomly drawn small samples of the local dataset and at the end we take the average of all this models. In this way, we are able to apply GreedyTL even to a big dataset. Moreover, this procedure can be easily parallelised on a multi core machine. As it will be clear in Section \ref{sec:perfEval}, this brings a strong impact on the generalisation performance\footnote{the performance of the algorithm applied to data not used during the training} of GreedyTL with respect to the performance of the local base model. 
%

\section{Distributed Learning procedures}
In this section we define  the distributed learning problem we cope with in this paper and the two distributed learning procedures we consider in our comparative study. 
\label{sec:distributed-learning}
\subsection {Problem description}
In this paper we consider the following scenario.  We assume that a dataset $D$ is distributed over different devices located in a number $L$ of different physical locations.  Therefore each location $l$ holds a partition $D_l, l=1,\dots,L$ of $D$. Clearly, we assume that the data contained in each partition $D_l$ are homogeneous, e.g., for a sensing application, they must be related to a well-defined physical entity that is sensed across different locations. This is necessary to meaningfully use, for the model corresponding to one partition, features of the models trained on the rest of the partitions.

In this paper, we assume that the partitions
$D_l$ are the training sets for training $l$ partial classifiers, which are then
refined based on each other features. Therefore, each local dataset is in the
form $D_l =  \{\bf{X}_l,Y_l\}$ where $X_l\in\mathbb{R}^{n_l\times d}$ is the set
of examples (or patterns) available at the $l$-th location $l$,
$Y_l\in\mathbb{R}^{n_l\times 1}$ is the set of labels that define the class to
which each pattern belongs to, $n_l$ is the cardinality of $D_l$ and $d$ is the
number of features of each pattern $\bf{x}\in\bf{X}$ (i.e., the dimensionality
of each data point in the dataset).  We consider that
all that nodes can communicate
 between each others through the network.
Specifically, instead of the data, they exchange the models they have
learnt from their local datasets.  Hereafter, according to the general
machine learning terminology, with the term machine we identify
  a computationally capable device that has direct access to the local dataset
in order to execute a learning procedure and learn a model of the data.
Specifically, a machine is therefore a machine learning algorithm running
at the node where the local dataset is stored. A model
is a parametric function $h$ that takes in input a pattern $\bs x \in \bs X$ and
returns in output a number $\hat y$ that corresponds to the class label to which
the pattern $\bs x$ belongs. The parameters of the function $h$ are the
quantities to be learnt by the machine learning algorithm.

\subsection{GTL: Distributed learning through Hypothesis Transfer learning}

Now we describe the distributed learning solution based on GreedyTL. It is
composed of 5 steps that are performed by each
machine. Algorithm~\ref{alg:DLGTL} shows the pseudo-code of this
algorithm.  

\paragraph*{Step 0 (Lines 1-5)} In the first step each machine at each location trains a learner on its local data. Although in principle every learning algorithm could be used during this phase, in our paper we use a Linear Support Vector Machine learning algorithm~\cite{Scholkopf:2001aa}. It is worth noting that it is possible to adapt the same procedure to other learning algorithms but we have chosen this learning algorithm because of its well known good performance and simplicity. 
After this first learning step, each location $l_i$ holds a model $h^{(0)}_{l_i}$ (Line \ref{alg:step1}).

\paragraph*{Step 1 (Lines 6-8)} After Step 0, there is the first synchronisation phase. All the models learnt at each location are sent to all the other locations. At the end of this phase each location holds $L$ SVM linear predictors, each one trained on different data. Note that at every location, models are stored in the same order, i.e. $h^{(0)}_{1}$ is the same model at each location. As it will be clear later on this is instrumental to enable the second step of model aggregation described in Step 4. 

\paragraph*{Step 2 (Line 9)} 
At each location we perform a second training phase on the same data using
GreedyTL. To this end we provide to it the set of models obtained during the
previous steps. The output of this step is the model defined in Equation
\ref{eq:htrg}. It is worth noting that the actual form of this model is
a single vector $\bs\omega' = [\beta^T,\omega^T]^T$ where the first $L$
components correspond to the $\bs\beta$ coefficients and the rest to the
$\bs\omega$ coefficients. Note that its dimensionality is greater than the
dimensionality of the problem, i.e. the problem has dimensionality $d$ while
$\bs \omega'$ has dimensionality $d+L$. However, as it will become clear in
Section~\ref{sec:noh} this does not represent an issue from the network traffic point of view because, thanks to the ability of selecting the most informative features of GreedyTL algorithm, the vector $\bs\omega'$ tends to be sparse.

\paragraph*{Step 3 (Lines 10-12)} After each location $l$ has its new model $h^{(2)}_{l}$, we have a new synchronisation phase in which, like for Step 1, every location sends its  $h^{(2)}_{l}$ model to all the other locations.

\paragraph*{Step 4 (Line 13)} Once all the models have been received, each location aggregates all of them into a single model $h^{(4)}$. As for Step 0, this aggregation can be done in several ways. Since we are framing into a classification problem, in this paper we adopted the majority voting and a consensus approach.  Majority Voting is a very common technique used to aggregate the predictions of many classifiers into one single response. Briefly, it selects the most frequent prediction as final response.  In the consensus aggregation, instead, $h^{(4)} = \frac{1}{L} \sum_{l=1}^{L} \bs\omega'_{l}$.  Of course these are just an example of the several and more sophisticated techniques one could apply at this Step. Here we used them because of their simplicity and, as we will see in Section \ref{sec:perfEval}, good performance. 
%
%

\begin{algorithm}[ht!]
    \caption{GTL Distributed Learning procedure \label{alg:DLGTL}}
    \begin{algorithmic}[1]
    \STATE Let be $L$ the number of data locations
    \STATE Let be $l_i$ the ID of the i-th location and $l_c$ the ID of the current location $c$.
    \STATE Let be $h^{(j)}$ the local model at step $j$
    \STATE Let be ${\bf X_c,y_c}$ the training patter and training labels for location $l_c$, respectively
    \STATE $h^{(0)}_{l_c} = TrainBaseLearner({\bf X_c,y_c})$ \label{alg:step1}
    \STATE SendModelToAll($h^{(0)}_{l_c}$)
    \STATE  $H^{src}$   = ReceiveBaseModels();
    \STATE $H^{src} \leftarrow H^{src} \cup \{h^{(0)}_{l_c}\}$
    \STATE $h^{(2)}_{l_c} = GreedyTL({\bf X_c,y_c},H^{src})$
    \STATE SendModelToAll($h^{(2)}_{l_c}$)
    \STATE $H^{gtl}$ = ReceiveGTLModels()
    \STATE $H^{gtl} \leftarrow H^{gtl} \cup \{h^{(1)}_{l_c}\}$
    \STATE $h^{(4)}$ = CombineModels($H^{gtl}$)
\end{algorithmic}
\end{algorithm}


\subsection{noHTL: Consensus and Majority Voting based distributed learning}
Let us now introduce an alternative distributed learning procedure
that we used in our comparative study with GTL. As a matter of fact, these
procedures are a subset of the steps composing the GTL procedure. The main
difference between GTL and noHTL is the absence of the second training phase
, i.e., Steps 2 and 3 of GreedyTL.
There is, moreover, a difference in the case when Majority Voting or
  Consensus-based aggregation is used. In the former case, at each location, the
  outcome is the most selected class by all the local models. Therefore, all
  models need to be available at all locations. From an algorithmic standpoint,
  therefore, noHTL with Majority Voting is exactly the same as GTL with the
  exclusion of steps 2 and 3, i.e., Lines 9-12 of Algorithm~\ref{alg:DLGTL}. On
  the other hand, in case of Consensus-based aggregation it is possible to optimise the
  algorithm to reduce the network overhead. Specifically, we assume that,
  instead of sending all models computed at Step 1 to all other nodes, they are
  sent to a unique node, which we call models collector. This node computes the
  average of the models, and sends this back to all the other nodes. Therefore,
  the steps of the algorithm become as follows, captured in
  Algorithm~\ref{alg:nohtl}.

\paragraph*{Step 0 (Lines 1-5)} In the first step each machine at each location trains a learner on its local data as in the GTL procedure. For the sake of comparison we used the same base learner as for GTL. 
\paragraph*{Step 1 (Line 6-8)} After Step 0 all the models learnt at each location are sent to one device that acts as models collector. It is not important which device is selected as models' collector. This is the first main difference with GTL. As we can see in the experimental part, this step has strong impact on the network traffic generated by this procedure. 
\paragraph*{Step 2 (Line 9)} 
The model collector computes the average model similarly to Step 4 of the GTL procedure: $h^{(2)} = \frac{1}{L} \sum_{l=1}^{L} h_l^{(0)}$.
\paragraph*{Step 3 (Lines 10-11)} The model collector sends the mean model $h^{(2)}$ to all the other devices

\begin{algorithm}[ht!]
    \caption{noHTL Distributed Learning procedure \label{alg:nohtl}}
    \begin{algorithmic}[1]
    \STATE Let be $L$ the number of data locations
    \STATE Let be $l_i$ the ID of the i-th location and $l_c$ the ID of the current location.
    \STATE Let be $h^{(j)}$ the local model at step $j$
    \STATE Let be ${\bf X,y}$ the training patter and training labels for location $l_c$, respectively
    \STATE $h^{(0)}_{l_c} = TrainBaseLearner({\bf X,y})$ \label{alg:nohtl_step1}
    \STATE SendModelToAll($h^{(0)}_{l_c}$)
    \STATE  $H^{src}$   = ReceiveBaseModels();
    \STATE $H^{src} \leftarrow H^{src} \cup \{h^{(0)}_{l_c}\}$
    \STATE $h^{(2)}_{l_c} =  \frac{1}{|H^{src}|} \sum_{l=1}^{|H^{src}|} h_l^{(0)}$.
    \STATE SendModelToAll($h^{(2)}_{l_c}$)
    \STATE $h^{(2)}$ = ReceiveModel()
\end{algorithmic}
\end{algorithm}

\section{Datasets description}
\label{sec:metrics}
In this section we present the datasets we used to evaluate the considered solutions. 
We used two well known datasets containing real world measurements, the first one, the HAPT dataset, is commonly used in the field of Human Activity Recognition (HAR))\cite{Reyes-Ortiz:2015aa}, while the second one is the well-known MNIST dataset that contains images of handwritten digits~\cite{LeCun:1998aa}. 
Hereafter we provide a detailed description of both datasets. 

\subsection{HAPT dataset} 

\begin{figure}[ht!]
\centering
    \includegraphics[width=.5\columnwidth]{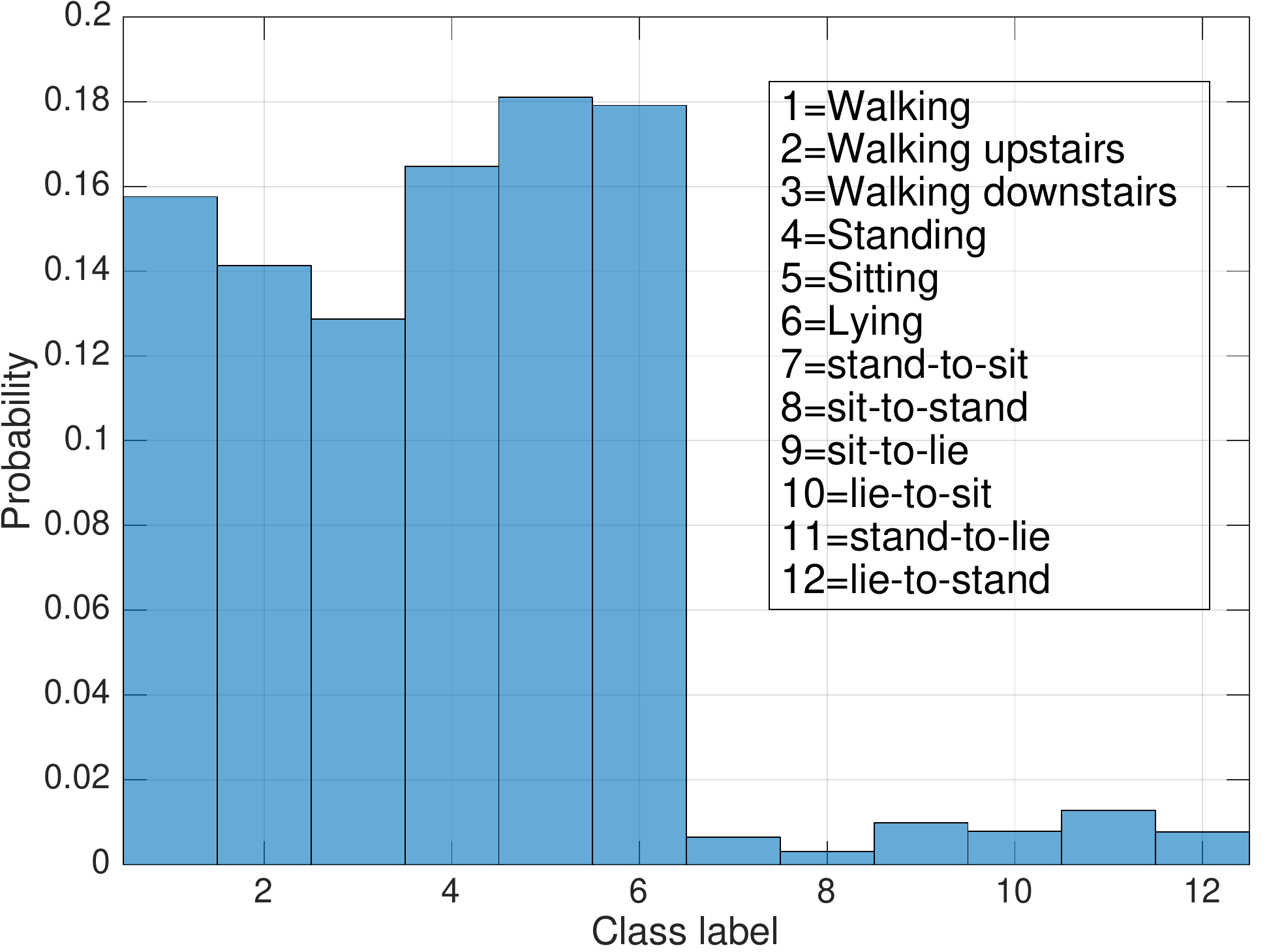}
    \caption{Probability density function of the activities performed by each user.\label{fig:dataDist}}
\end{figure}

The dataset was built during experiments carried out with a group of 30 volunteers within an age range of 19-48 years. 
Each volunteer performed a protocol of activities composed by six basic activities: three static postures (standing, sitting, lying) and three dynamic activities (walking, walking downstairs and walking upstairs). In the dataset are present also postural transitions that occurred between the static postures (e.g. stand-to-sit, sit-to-stand, sit-to-lie, lie-to-sit, stand-to-lie, and lie-to-stand). 
All the participants were wearing a smart-phone (Samsung Galaxy S II) on the waist during the experiment execution. 
The raw measurements are 3-axial linear acceleration and 3-axial angular velocity at a constant rate of 50Hz using the embedded accelerometer and gyroscope of the device. 
Afterwards, the sensor signals (accelerometer and gyroscope) were pre-processed by applying noise filters and then sampled in fixed-width sliding windows of $2.56$ sec and $50\%$ overlap ($128$ readings/window). 
From each window, a vector of $561$ features was obtained by calculating variables from the time and frequency domain.
The resulting dataset is composed by $N=10929$ records where each one is a vector $x\in\mathbb{R}^d, d=561$. The total size in MB of the raw measurements is 103MB that is greater than the size of the post-processed dataset (48MB). 
It is worth noting that, the distribution of activities for each user is skewed:
static and dynamic postures are performed more frequently than transitions, as
reported in Figure~\ref{fig:dataDist}. 
Moreover,  $9$ local datasets out  of $30$ do not have a complete set of
activities, therefore in our experiments we consider only those users
that present a complete set of activities. The data corresponding to the
excluded users have been uniformly distributed between the remaining $21$
users. Note that the dataset is unbalanced, in the sense that different
  classes have quite different number of samples on which the models can be
  trained. As this is a critical point, we study in detail the behaviour of
  distributed learning in cases of various types of unbalanced datasets in
  Section~\ref{sec:peva}.

\subsection{MNIST dataset}
MNIST is a database of $70000$ handwritten digits.  
Images contained in the dataset are numbers from 0 to 9.  Digits have been normalized and centered in a fixed size 28x28 image, meaning that each image in the dataset corresponds to a record of 784 raw features. 
As it is common practice when working with images, we transformed each one into a more informative feature set. Specifically, we transformed each image into a histogram oriented graph (HOG) \cite{Dalal:2005aa}.  The image is divided into a grid of small cells, and for the pixels within each cell, a histogram of gradient directions is computed, i.e. gradients are computed on the gray-scale value of pixels. The resulting HOG descriptor for an image is the concatenation of these histograms.
In our experiments we used a cells grid of size $5\times 5$, meaning that the image is divided in 25 regions over which histograms are computed. 
After applying the HOG feature extraction method, each image is transformed into a feature vector of size $324$.
Note that this procedure is applied to each image separately therefore this workflow is perfectly compatible with a crowd sensing scenario like the one we are considering, i.e. pictures taken by users are preprocessed before being processed by our distributed learning system.

\begin{figure}[H]
    \centering
    \subfloat[]{
    \includegraphics[width=.35\textwidth]{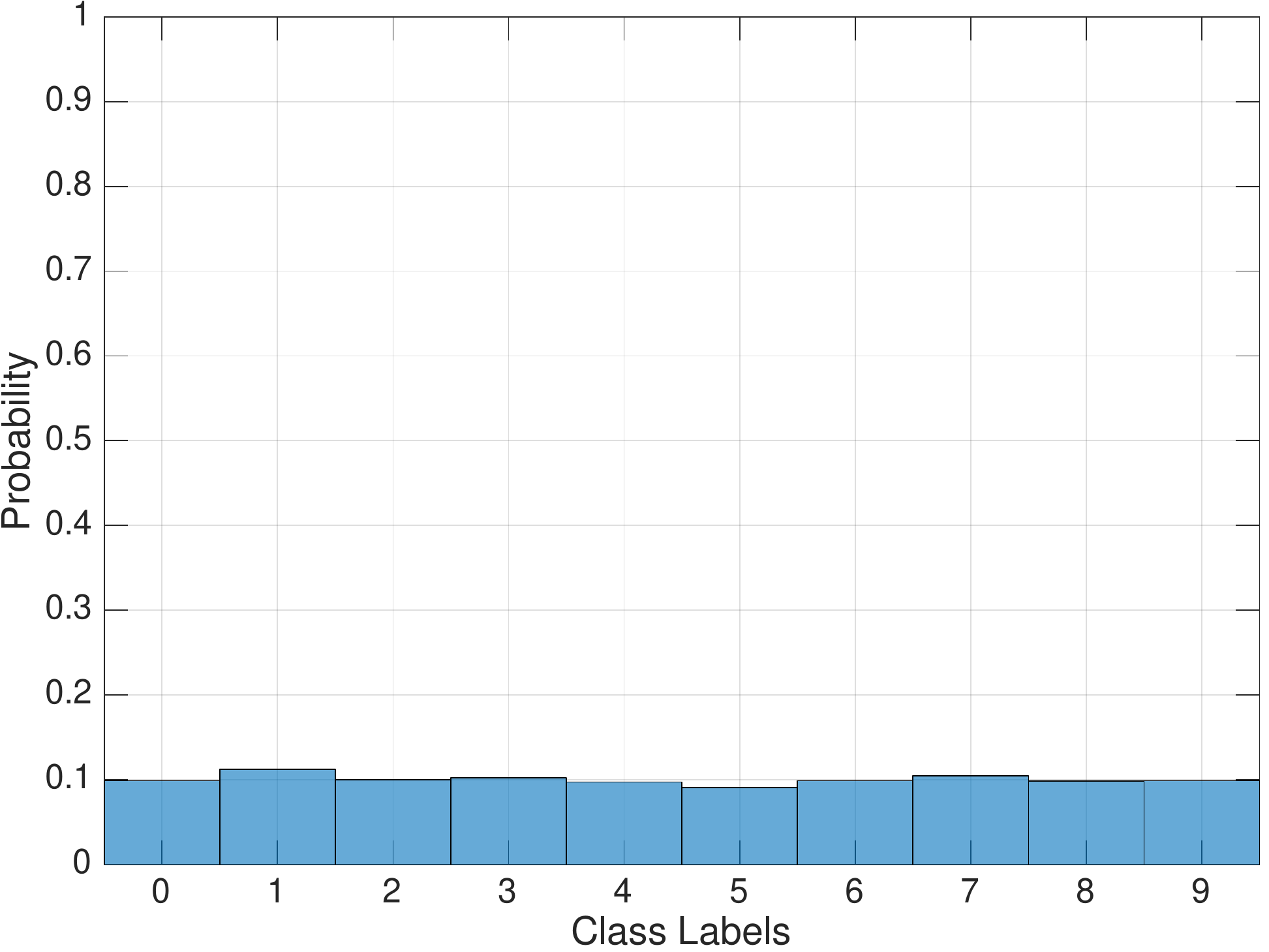}
    \label{fig:mnist_balanced}
}
    \subfloat[]{
    \includegraphics[width=.35\textwidth]{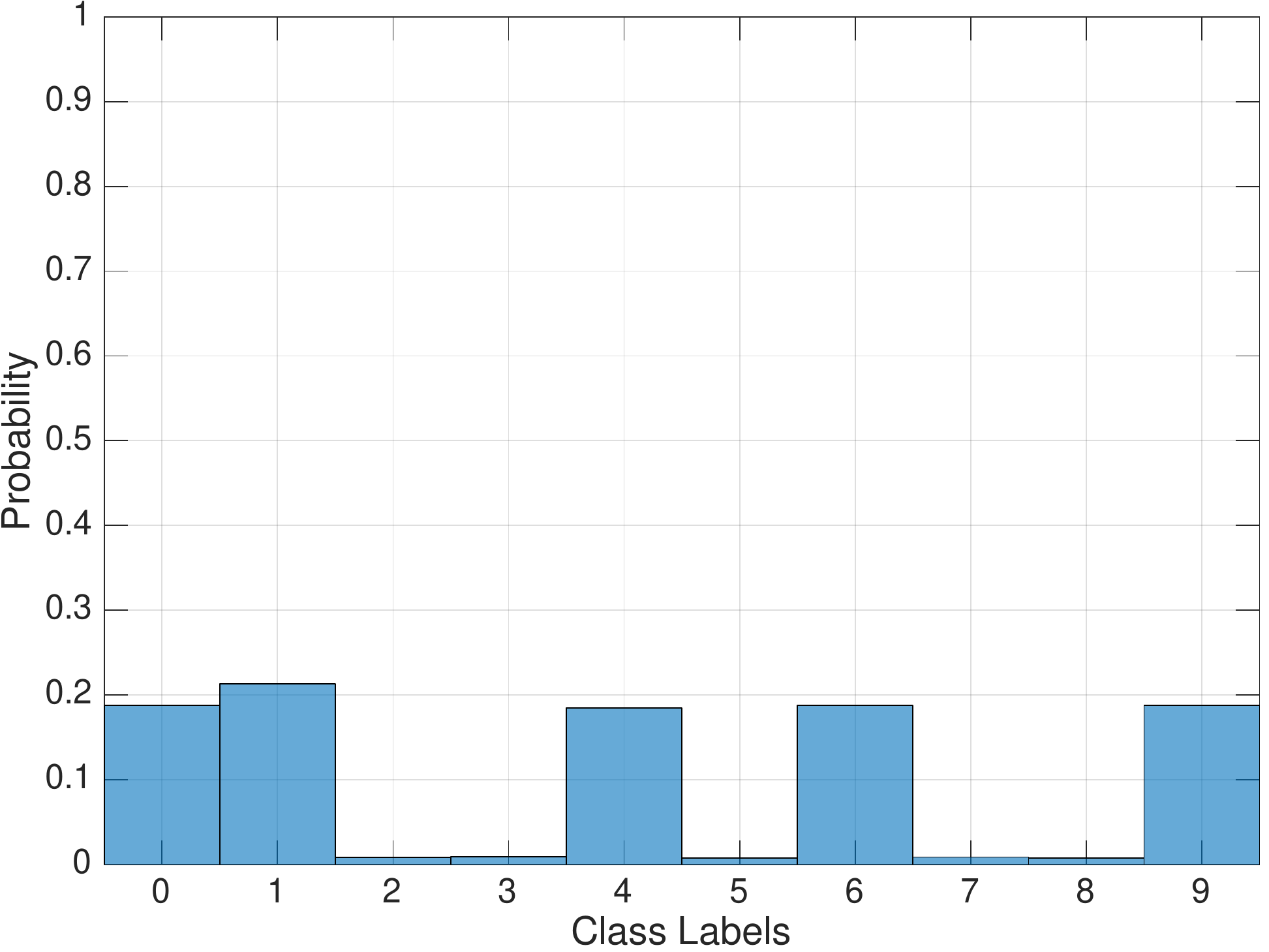}
    \label{fig:mnist_36789}
}\\
    \subfloat[User 1]{
    \includegraphics[width=.35\textwidth]{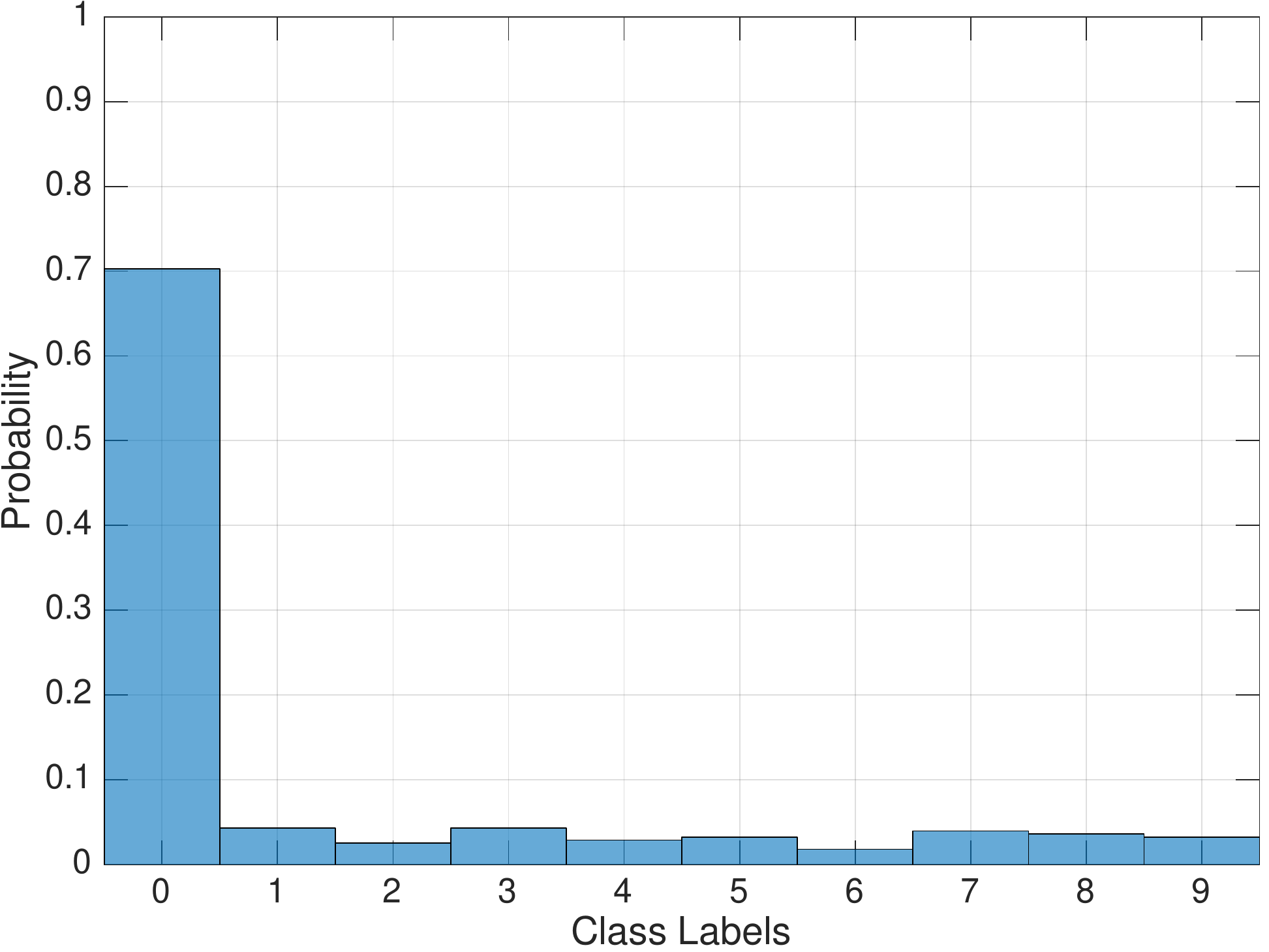}
    \label{fig:onehot1}
}
    \subfloat[User 2]{
    \includegraphics[width=.35\textwidth]{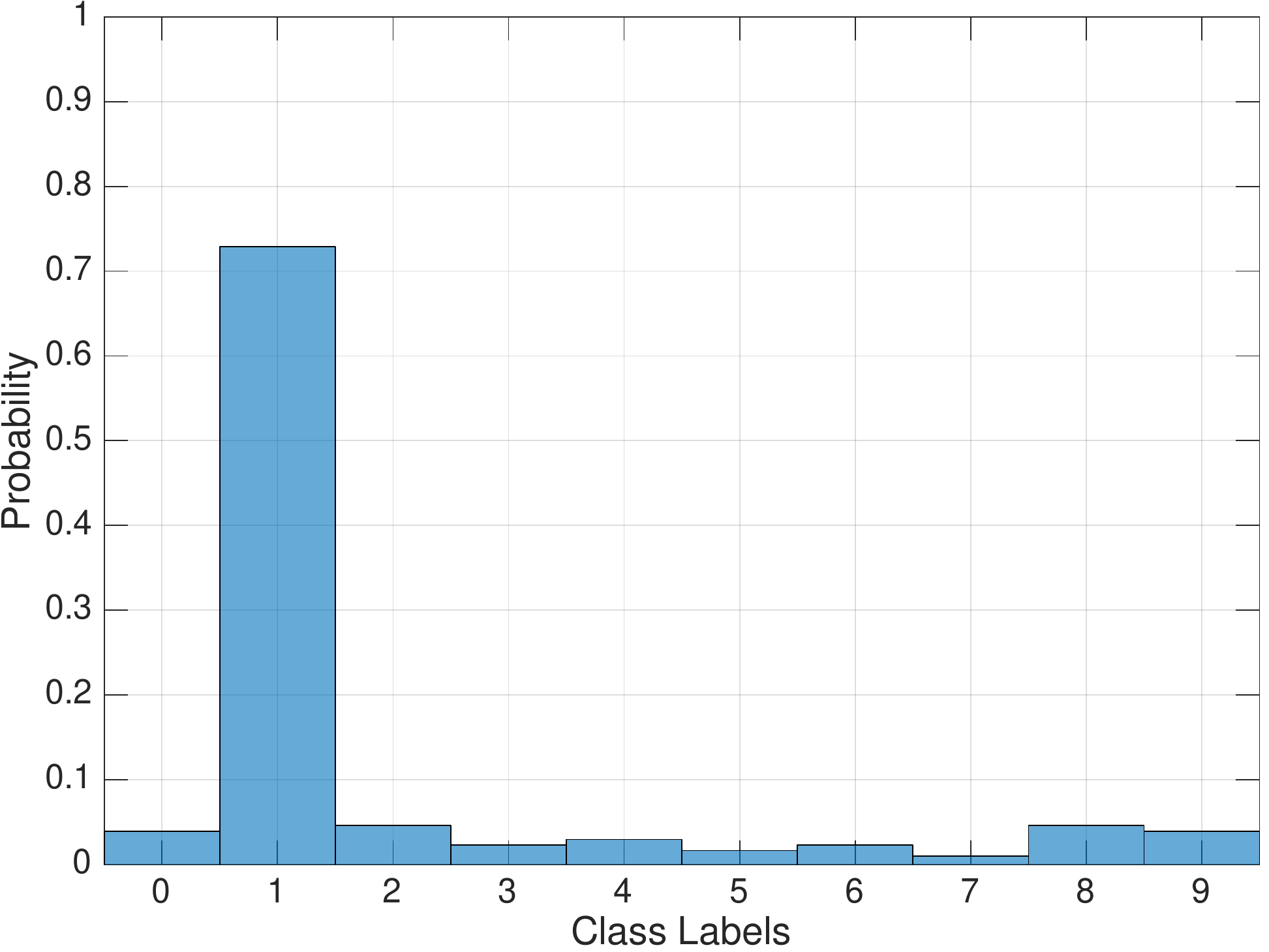}
    \label{fig:onehot2}
}
\caption{Per user class distribution for each MNIST-based dataset.}
\end{figure}

Similarly to the HAR scenario, we consider to have 30 users involved in a mobile
crowdsourcing task in which each of them collects a certain number of images on
their phones. In this paper we considered three possible scenarios. First, the
classes distribution of digits per user is uniform, i.e. each user collected
almost the same number of images per class (see Figure~\ref{fig:mnist_balanced}). 
Second, the distribution of classes per user is equally skewed for all the
users, i.e. for each device the classes of digits 2,5,6,7,8 are less represented
then the others (see Figure \ref{fig:mnist_36789}). Third, the classes
distribution between users is different. Precisely, we simulated a ``one hot''
situation in which each user's device collected many examples of one class and
only few for the rest of the classes (see Figure
\ref{fig:onehot1}-\ref{fig:onehot2}). The most popular class is rotated
among the 10 possible classes among the users, and thus each class is the most
popular for 3 users.

The modifications in the second and third cases allow us to study a very
  important property of our proposed approach. Using these modifications of the MNIST dataset we are able
to evaluate the performance of the proposed approach by unbalancing the dataset
in a controlled way. Specifically, in the second case we consider unbalancing
between classes, but uniformity across locations, i.e., the number of data points
are different across classes, but all locations ``see'' the same distribution of
data points. This type of unbalancing (that is present also in the original HAPT
dataset) is throughout referred to as ``class unbalance''.
In the third case the dataset is balanced across classes, but unbalanced across
nodes, as different nodes ``see'' a different distribution of data points for
training. This type of unbalance is throughout referred to as ``node
unbalance''. Note that one of the goals of GTL is to overcome unbalance of
both types, by (i) transferring and refining knowledge obtained over
limited-size datasets in case of class unbalance, and (ii) transferring
knowledge acquired at locations with ``more information'' to locations with
``less information'' in case of node unbalance`. We anticipate that, as shown in
Section~\ref{sec:peva}, our GTL-based algorithm is able to achieve this goal.

\section{Results on prediction performance }
\label{sec:peva}

We evaluate the performance of distributed learning using the two datasets
  in Sections~\ref{sub:caseI} and \ref{sub:caseII}, respectively.
  Sections~\ref{sub:skewed} and \ref{sub:one-hot} considers the case of MNIST with skewed
  distributions of samples, described in Section~\ref{sec:metrics}. 
 
  We recall
that we compare the performance of GTL and noHTL to characterise the prediction
performance vs. network overhead trade-off. We also compare them with a
centralised solution where all data is sent to a central cloud (this approach is
hereafter referred to as ``Cloud''). Before presenting the results,
in Section~\ref{sec:perfEval}, we discussed the performance indices used in the
analysis.

\subsection{Performance metrics}
\label{sec:perfEval}
In this paper we evaluate the prediction performance  of our system using the well known F-measure~\cite{Powers:2011aa}.
The F-measure is a common aggregate index mostly used to evaluate the
performance of a content retrieval system or a classifier. We use it because it
provides a more precise description of the performance of our system with
respect to the simple accuracy index defined as the number of correct
predictions divided by the total number of predictions made. Precisely, through
the F-measure we are able to summarize in one single index the performance
expressed by the Precision and Recall indexes. The F-measure is defined as the harmonic mean between the precision index and the recall index, hereafter defined. Note that in this paper all the performance indexes are location-wise, i.e. they always refer to performance obtained by models trained in each separate location $l$.

The precision (or specificity) index is  defined as the number of correct predictions divided by the total number of predictions made. More formally: 
\begin{equation}
    p_l =  \frac{1}{m_l}\sum_{i=1}^{m_l} I(y_i,\hat{y_i})
\end{equation}
with 
$$I(y_i,\hat{y_i})=\left\{\begin{array}{c c} 1 & \mathrm{if\ } y_i=\hat{y_i} \\ 0 & \mathrm{otherwise}\end{array}\right. $$ where $\hat{y}$ is the predicted class of the i-th pattern $x_i$ and $y_i$ is its true class and $m$ is the number of the test set.  

The recall (or sensitivity) index is defined as the number of correct predictions for a specific class divided by the number of the patterns that belong to that specific class, averaged over all classes. More formally: 
\begin{equation}
    r_l = 
    \sum_{\forall c \in C}\frac{1}{m_{l,c}}\sum_{i=1}^{m_{l,c}}I(y_{c,i},\hat{y}_{c,i})
\end{equation}
Finally the F-measure is defined as follows:
\begin{equation}
    F_l = 2*\frac{p_l \cdot r_l}{p_l+r_l}
\end{equation}
This measure takes values in the range $[0,1]$, where $0$ stands for the worst prediction performance and $1$ to the best one. 

We define the Prediction Performance Gain (PPG) index as the percentage increment of prediction performance we gain after each step
of both GTL and noHTL procedures, against the performance of the local base
classifier, i.e. the local model learnt after Step 0.
More formally, let us denote with $F_l^{h^{(0)}}$ the F-measure value  of the
local model trained at Step 0 for location $l$ and with $F_l^{h^{(j)}}$ with
$j>1$ the model obtained at the $j$-th step. We compute the PPG gain
$\rho_l^{(j)}$ at location $l$ and step $j$ as

\begin{equation}
  \rho_l^{(j)} =   1-\frac{1-F_{l}^{h^{(j)}}}{1-F_{l}^{h^{(0)}}}
\end{equation}

As, in general, $1-F$ is 1 when the classifier performs at worst,
  $\rho_l^{(j)}$ ranges from 0 and 1, and tends to 1 the more the various steps
  (specifically, steps
  2 and 4 for GTL and noHTL with Majority Voting, Step 2 for noHTL with
  Consensus-based aggregation)
improve performance with respect to step 0. Conversely, if  $\rho_l$
  assumes a negative value, it means that  the models obtained at the
  various steps are less accurate of the local base model.

We tested our framework adopting a ``70-30'' hold out procedure. Precisely,
every round we keep the $30\%$ of data as test set,  and we apply our distributed
learning procedure on the remaining $70\%$. Then the performance results are
averaged over 10 runs and confidence intervals at $95\%$ are computed. 
Note that for each run the data contained in the test set is never presented to learning algorithm during the training phase. We use it only to evaluate the performance of the  model learnt at the end of the training phase.

Note that, the problem at hand falls in the domain of  multi-class
classification. The approach adopted in this paper to cope with more than two
classes is the well-known ``one-vs-all''. Specifically, here each multi-class
classifier is composed by $k$ binary classifiers each one trained on its own
class. For example, for the binary classifier corresponding to class 1, all the
examples belonging to class 1 are treated as positive examples (labelled as
$+1$) and the rest are negative examples (labelled as $-1$).

In order to decode the final response of the multi-class classifier we adopt the
following standard approach. To each class label we associate a
binary  string of size $k$, e.g. if $k=3$, the string associated to the class
label $c=2$ is $b_2 = \{-1,1,-1\}$.  When a pattern $\bs x$ is presented to all
the $k$ trained binary learners, each of them provides its classification
response ($+1$ or $-1$).  Following the same example, all the classifiers'
responses form a binary string $\hat{b}$  , e.g.  $\hat{b}=\{-1,1,-1\}$ which,
in this particular case means that the second classifier says that the pattern
belongs to his own class, while the others say that the pattern does not belong
to their own classes.  The final response of the classifier corresponds to the
class label corresponding to the binary string $b_c$ that is more similar to the
response string $\hat{b}$. More precisely:   $$\hat{y} =
\argmin\limits_{c\in\{1,\dots,k\}}\sum_{i=1}^k \max(0,1- \hat{b}[i] \cdot
b_{c}[i])$$  where $\hat{y}$ is the final response of the multi-class
classifier, and $b[i]$ denotes the $i$-th element of string $b$. 


\subsection{Case study I: Human Activity Recognition (HAPT dataset)}
\label{sub:caseI}

Let us now present the performance related to the Activity Recognition
dataset. We point out that the confidence intervals we obtained for all the results presented in this and the following Sections are very tight, therefore, we do not show them for the clarity of the plots.

\begin{figure}[ht!]
\centering
\subfloat[]{
\includegraphics[width=.48\columnwidth]{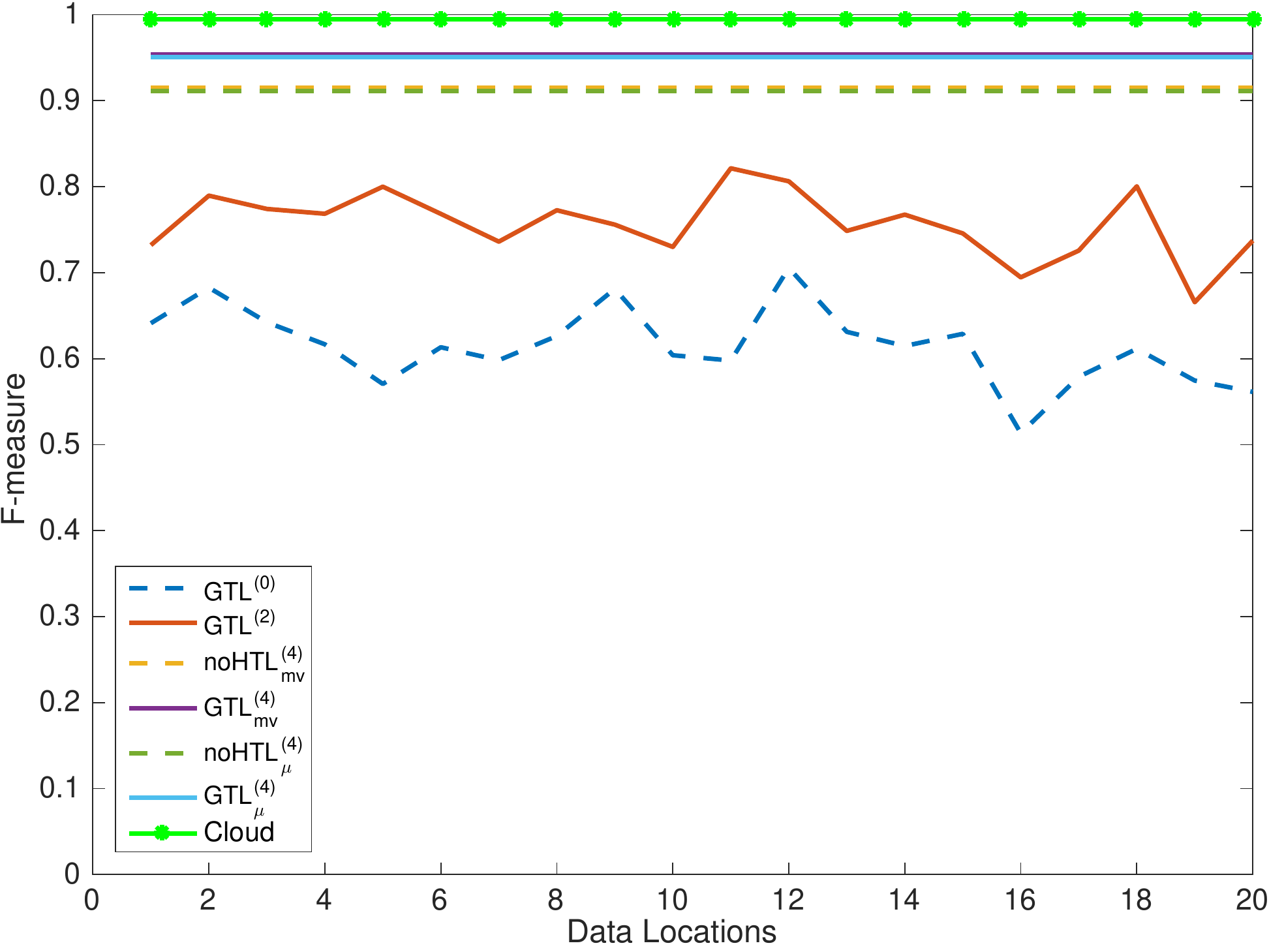}
\label{fig:acc}
}
\subfloat[]{
    \includegraphics[width=.48\columnwidth]{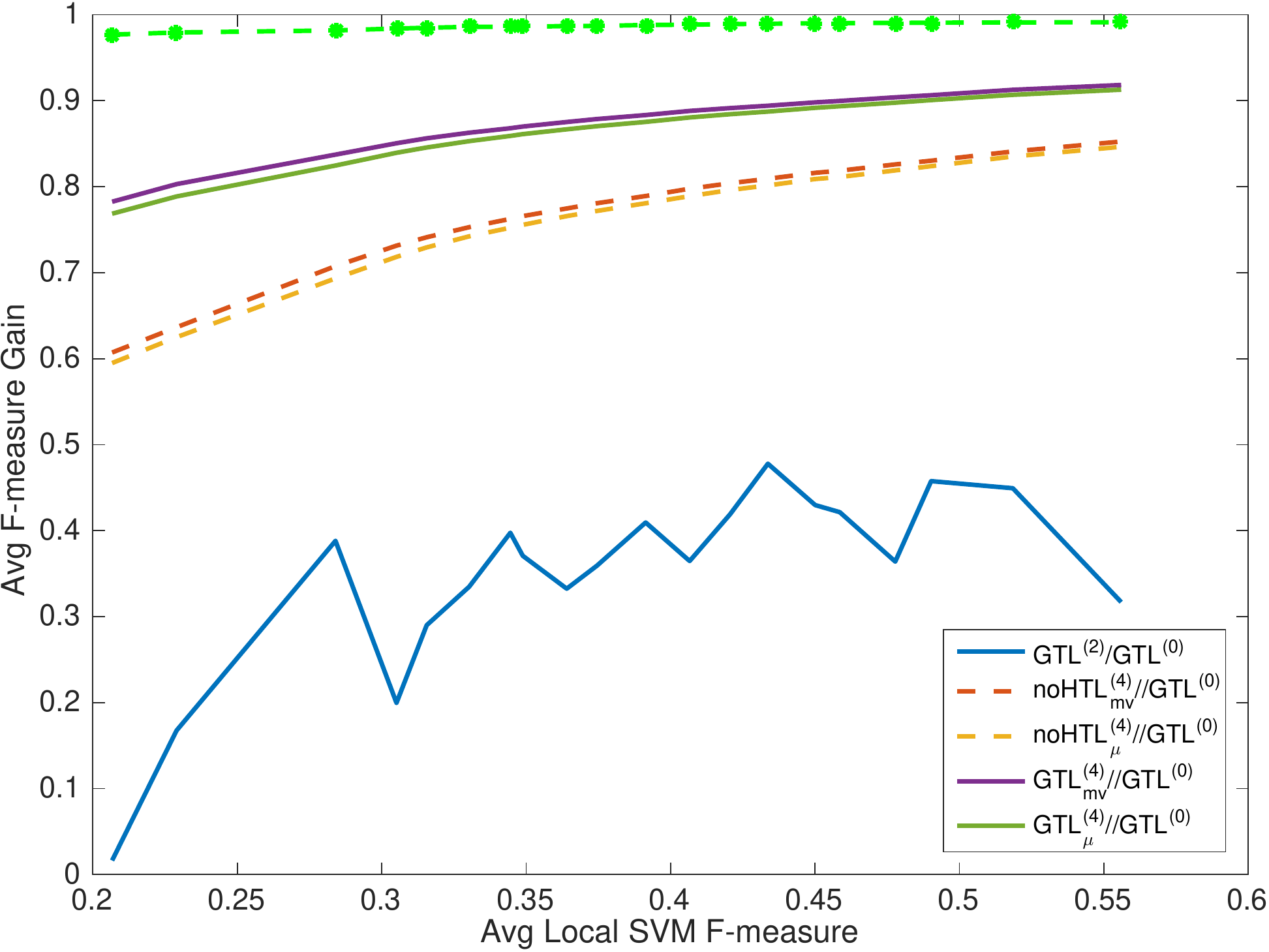}
\label{fig:gain}
}
\caption{HAPT dataset. Prediction performance (a) and Prediction Performance Gain (b) comparison per single location after each step  of the proposed distributed learning approach for both GTL and noHTL versions.}
\end{figure}

Figure \ref{fig:acc} shows the F-measure, in each data location, after each step of the proposed procedure.
As one can expect, the local models trained at the Step $0$ (here denoted for
simplicity by GTL$^{(0)}$ but also valid in the
noHTL$^{(0)}$, since both procedures have the same step 0), yield the worst
performance. This is due to the fact that, although SVM is a very good learning
algorithm, in the local data there are not enough examples to learn a model with
good generalisation abilities.
After the models' exchange at Step $1$, we can see that the HTL algorithm is able to well exploit the knowledge learnt at each location and learn a model with better performance (GTL$^{(2)}$). Note that each  GTL$^{(2)}$ model has been trained on the same local data on which was trained the GTL$^{(0)}$ models.
After the second exchange of models, the performance of GTL are even better. Note that after the second exchange of models there are no more differences in terms of performance between the locations. This is due to the fact that at each location  all the GTL$^{(2)}$ models learnt in all the other locations  are aggregated into a single model (GTL$^{(4)}$), which is the same for all locations.   
In fact, looking at Figure~\ref{fig:acc}, we can see that the performance of
the aggregated models based on HTL ($\mu$-GTL$^{(4)}$ and $mv$-GTL$^{(4)}$ stand
for mean model and majority voting, respectively) get closer and closer to the
performance obtained with a Cloud solution. In fact the F-measure with GTL is $0.95$ while the  one with the cloud approach is $0.995$.
As far as noHTL is concerned, we notice that in this specific  case its
performance in both  variants is better than  GTL$^{(0)}$, meaning that the
model aggregation procedure is beneficial to improve the generalisation
performance of each device. However, noHTL is not able to reach the GTL
performance. As we see in the next set of results, this is due
to the fact that noHTL is not able to compensate
for insufficient representation of some classes (remember that the HAPT
dataset is ``class unbalanced''). 
A third interesting fact is that both $mv-$noHTL and $\mu$-noHTL
have the same performance, reaching a final F-measure  of $0.92$. 

In Figure~\ref{fig:gain} we show the performance gain of each step w.r.t. the
performance of the local model. Specifically, for each step and each location we
compute the corresponding performance gain. We sort locations in
increasing order of loss of the local model GTL$^{(0)}$,
and use this ordering in the $x$ axis. In other words, the left-most values in
the graph correspond to the gains obtained in the locations with the lowest
F-measure of the base learner. Interestingly, the ranking shown in Figure \ref{fig:acc} is preserved. In fact
we notice that both GTL and noHTL improve the performance w.r.t. the
local models. However in this scenario it is clear that
simply averaging the local models is not enough to obtain good
performance.
It is also interesting to note that the gain of GTL solutions increases for
locations where the local model performs worse. This means
that, besides always improving over the local model, the
HTL-based solution ``helps more'' the locations with a lower initial
performance, which is also a desirable feature.

\begin{figure}
\centering
\subfloat[]{
    \includegraphics[width=.48\columnwidth]{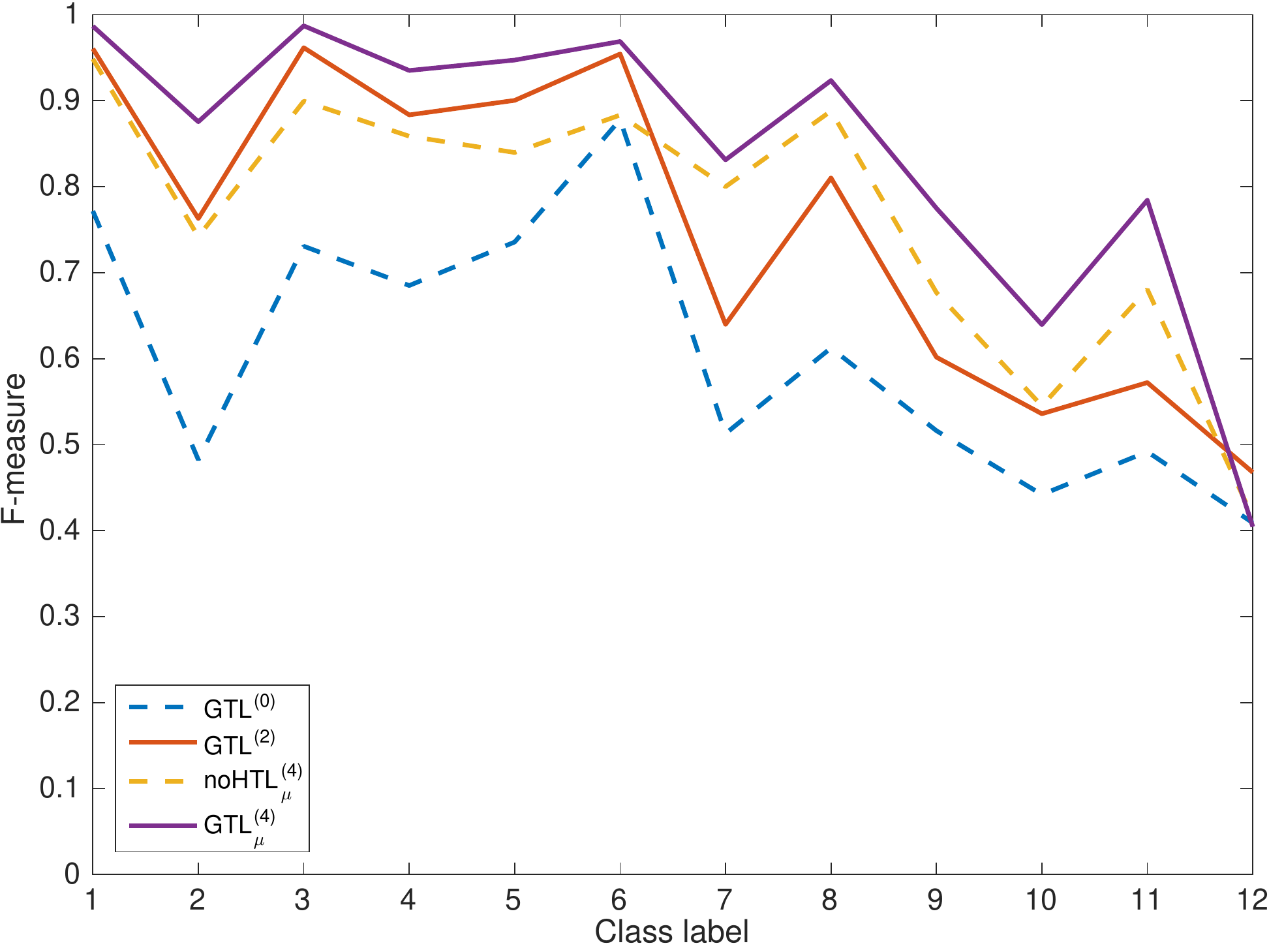}
    \label{fig:classAcc}
}
\subfloat[]{
    \includegraphics[width=.48\columnwidth]{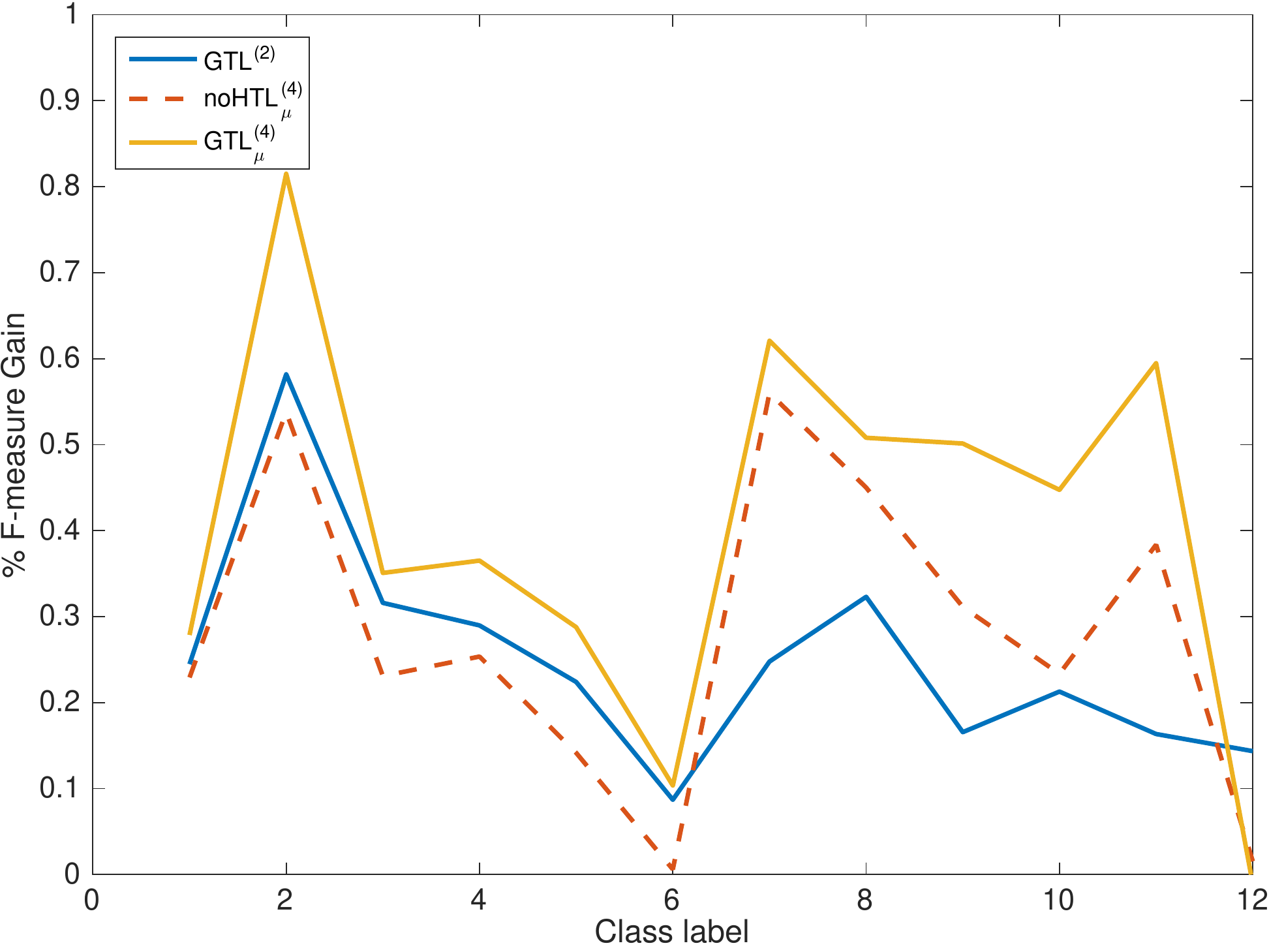}
    \label{fig:classAccGain}
}
\caption{(a) HAPT dataset. Prediction performance per class before and after the first and the second models' exchange. (b) Prediction performance gain per class after the first and the second model exchange.}
\end{figure}

In order to understand the real benefit coming from the use of the considered 
distributed learning approaches, we analysed the accuracy per single class,
averaged over all the locations for each step of each procedure. In this
way we want to highlight that even though the labels' distribution is
unbalanced, there is an advantage in averaging local models (as both GTL
and noHTL do), and that it is even more beneficial to transfer knowledge, as GTL
does and noHTL does not. In fact,
looking at Figure~\ref{fig:classAcc} we can notice that the unbalanced dataset
strongly affects the performance of GTL$^{(0)}$ (remember from
  Figure~\ref{fig:dataDist} that classes 1 to 6 are way more represented in the
dataset with respect to classes 7 to 12).
With GTL$^{(2)}$, noHTL, and GTL$^{(4)}$
instead we notice an better performance. One reason behind it is
  that, in general, averaging local models reduces uncertainty and thus results
  in better performance. However, we clearly notice a difference in performance
  between GTL$^{(4)}$ and noHTL, due to the fact that GTL transfers knowledge
across nodes. By doing so, it is
able to aggregate knowledge obtained at individual locations over small
datasets (for specific classes). This is supported by Figure~
\ref{fig:classAccGain}, which shows that GTL obtains a
performance gain up to $80\%$. We point out that for the sake of clarity, we do
not report the values for  $mv$-GTL$^{(4)}$ because they are equal to the ones
of $\mu$-GTL$^{(4)}$.

\subsection{Case Study II: Unmodified MNIST Dataset}
\label{sub:caseII}

\begin{figure}[ht!]
\centering
\subfloat[]{
\includegraphics[width=.48\columnwidth]{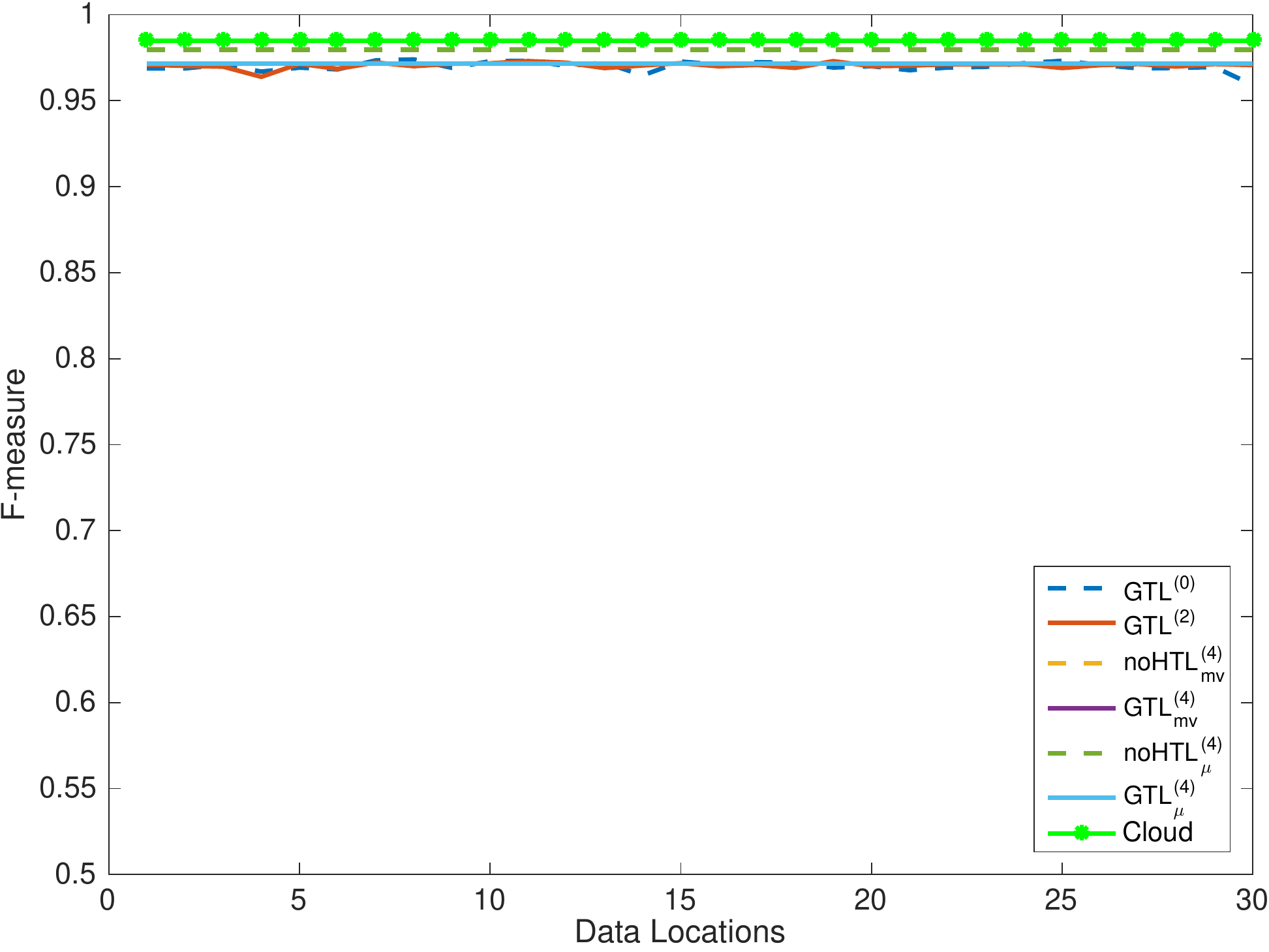}
\label{fig:mnist_bal_acc}
}
\subfloat[]{
\includegraphics[width=.48\columnwidth]{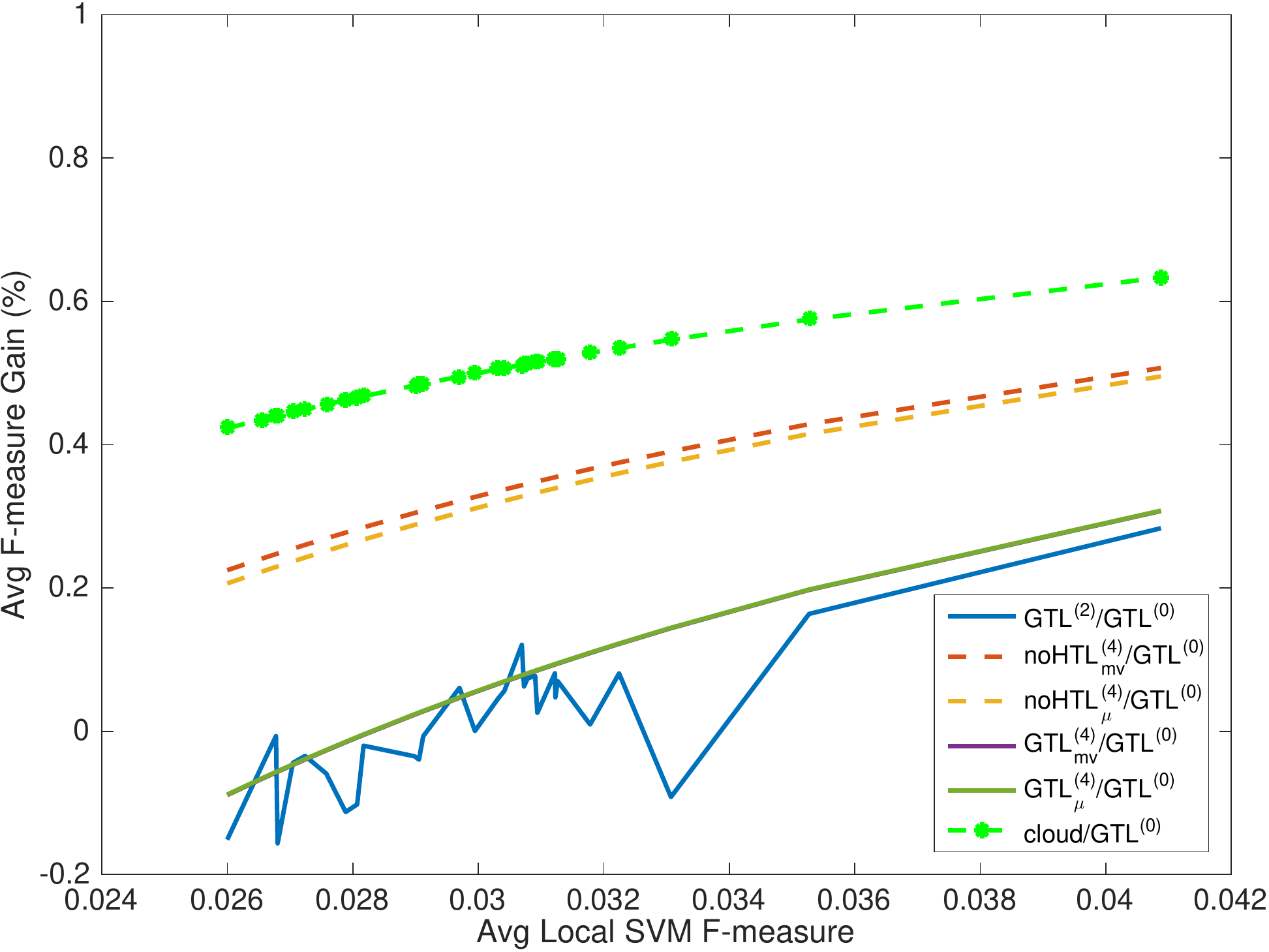}
\label{fig:mnist_bal_gain}
}
\caption{MNIST balanced. Prediction performance (a) and prediction performance gain (b) comparison per single location after each step  of the considered distributed learning approaches for both HTL and noHTL versions.}
\end{figure}

In the case of the unmodified MNIST dataset, all location see the same
distribution of data points per class, and all classes are equally represented.
Therefore, there is no unbalance of any kind. Specifically, the distribution of classes in
each local dataset owned by each device is of the form represented in Figure
\ref{fig:mnist_balanced}.

Looking at Figures \ref{fig:mnist_bal_acc}-\ref{fig:mnist_bal_gain} we notice
that both  noHTL  and  GTL 
yield very good performance, quite close to the one
of the Cloud solution. As a matter of fact, in this case even local models are already
quite good, due to the fact that classes are fairly well represented at each
location. When comparing GTL and noHTL, it is interesting
  to note that noHTL outperforms GTL. The reason is that, as the dataset is
  totally balanced, aggregating local models (as noHTL does) is sufficient,
  while transferring knowledge may lead to over fitting, and sometimes even
  deteriorates performance with respect to local models (see
  Figure~\ref{fig:mnist_bal_gain}). However, remember that our goal is not to
  champion GTL in all cases, but to show that distributed learning can be as
efficient as cloud-based approaches. In this specific case, the best distributed
learning solution is noHTL instead of GTL.

\begin{figure}[ht!]
\centering
\subfloat[]{
    \includegraphics[width=.48\columnwidth]{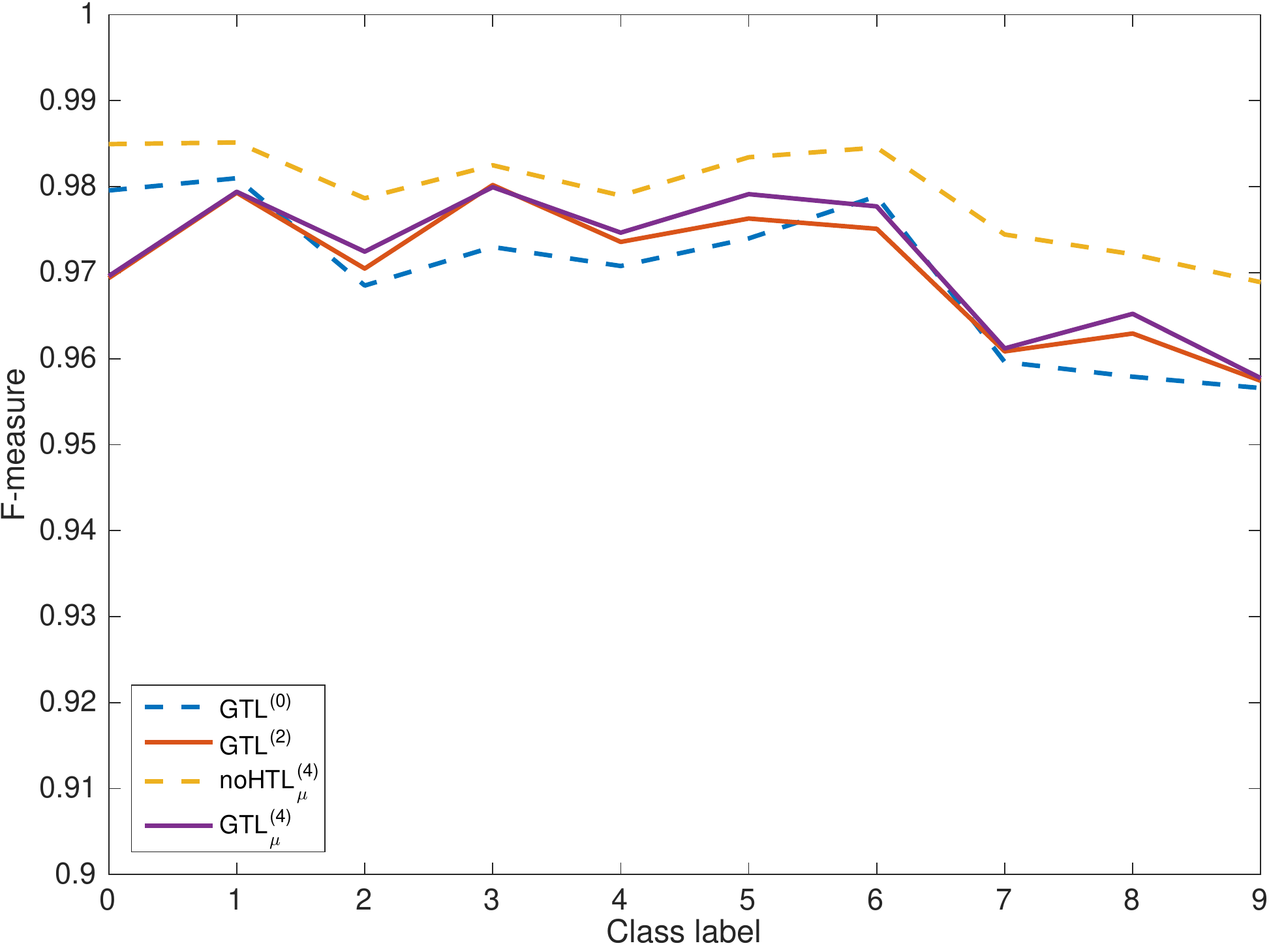}
    \label{fig:mnist_bal_classAcc}
}
\subfloat[]{
    \includegraphics[width=.48\columnwidth]{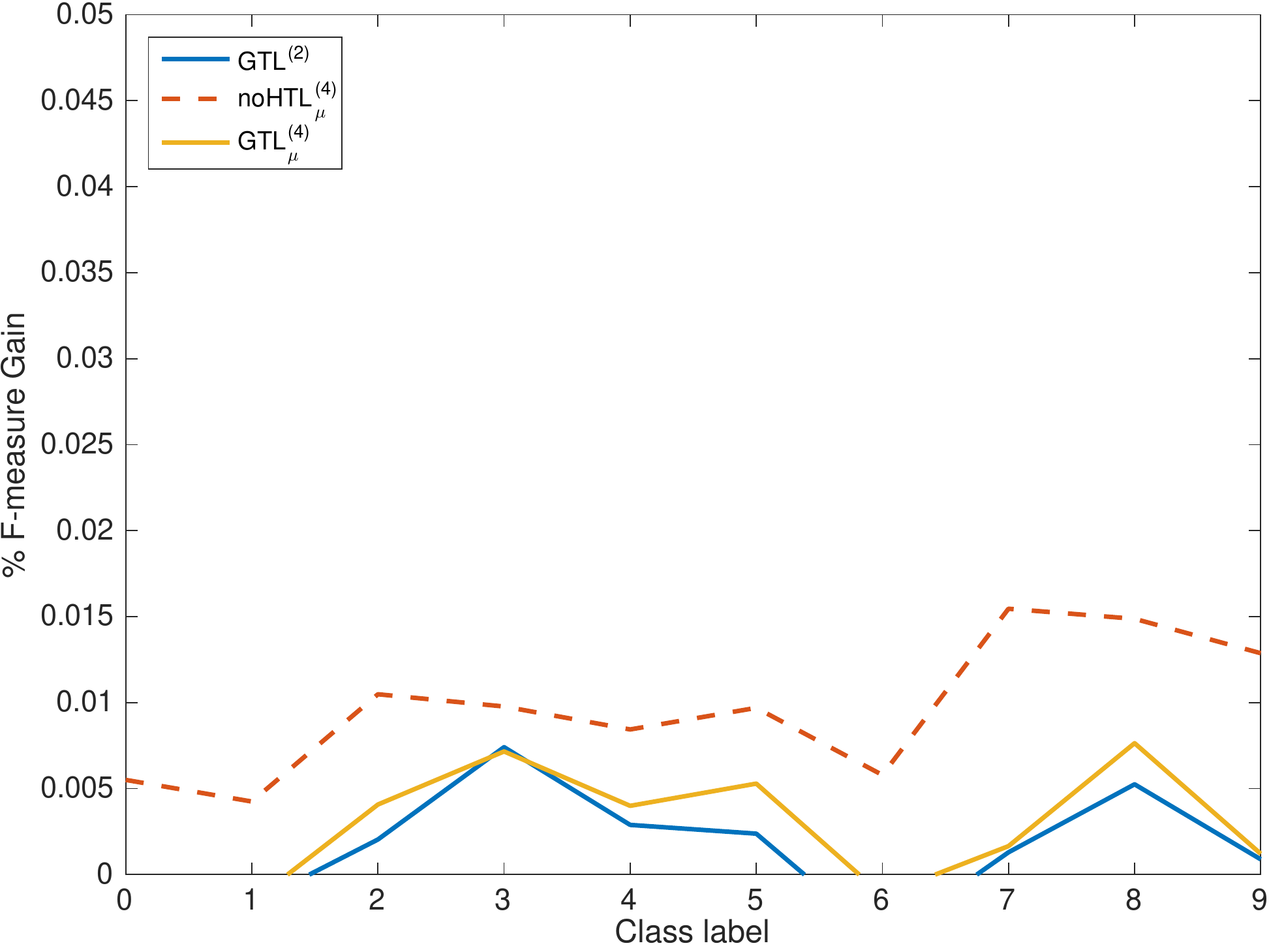}
    \label{fig:mnist_bal_classAccGain}
}
\caption{MNIST balanced. (a) Prediction performance per class before and after the first and the second models' exchange. (b) prediction performance gain per class after the first and the second model exchange.}
\end{figure}

Figures~\ref{fig:mnist_bal_classAcc} and \ref{fig:mnist_bal_classAccGain}
show the performance on a per-class basis. This analysis confirms the initial
findings. In this case, as the dataset is balanced and all classes are already
well represented at all locations, the advantage of transferring knowledge or
aggregating local models is minimal, if any. Local models already yield an
F-measure close to 1, as in the case of Cloud (but, clearly, without generating
\emph{any} traffic on the network).

\subsection{Case Study III: MNIST with class unbalance}
\label{sub:skewed}
\begin{figure}[ht!]
\centering
\subfloat[]{
\includegraphics[width=.48\columnwidth]{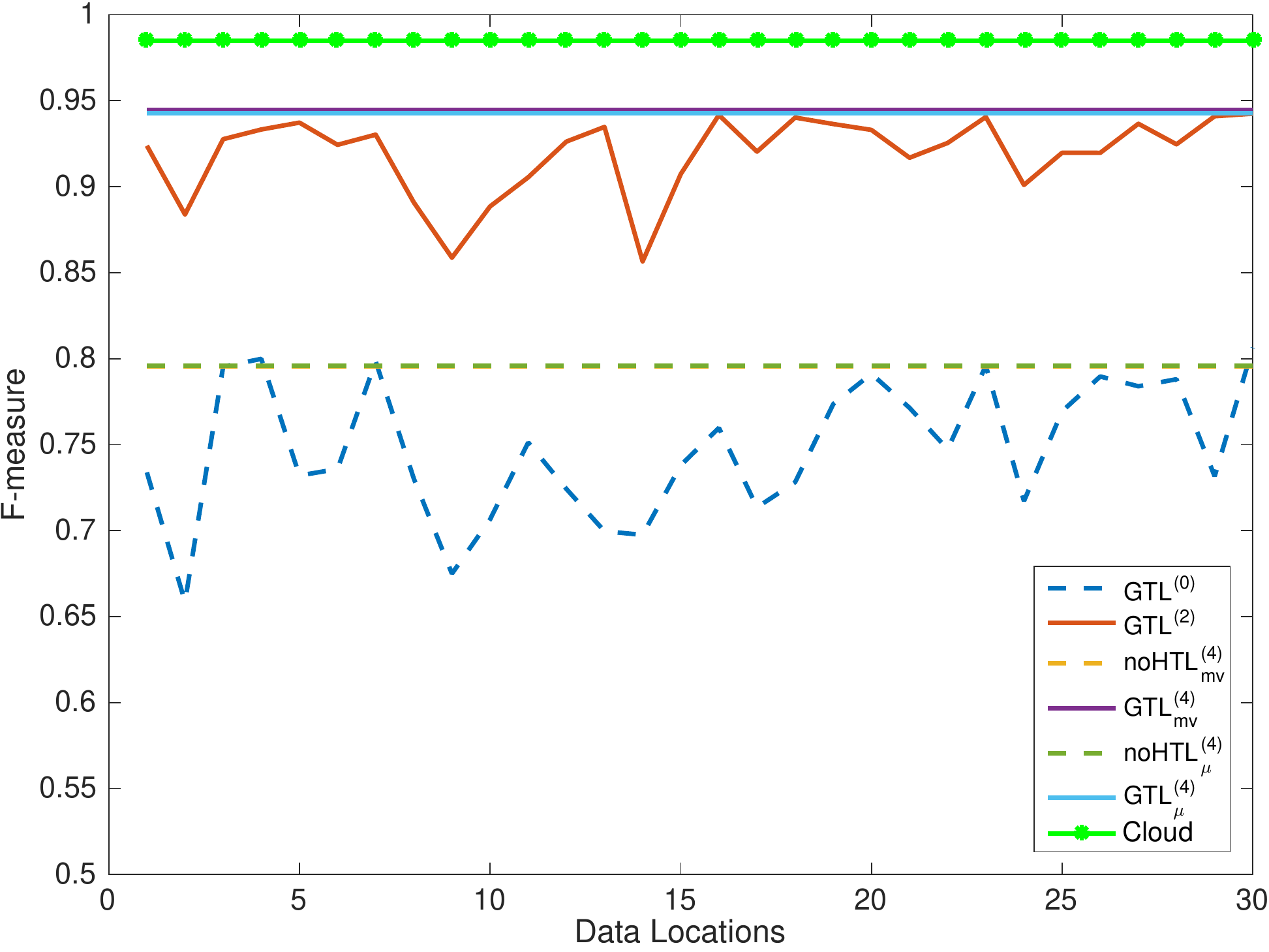}
\label{fig:mnist_unbal_acc}
}
\subfloat[]{
\includegraphics[width=.48\columnwidth]{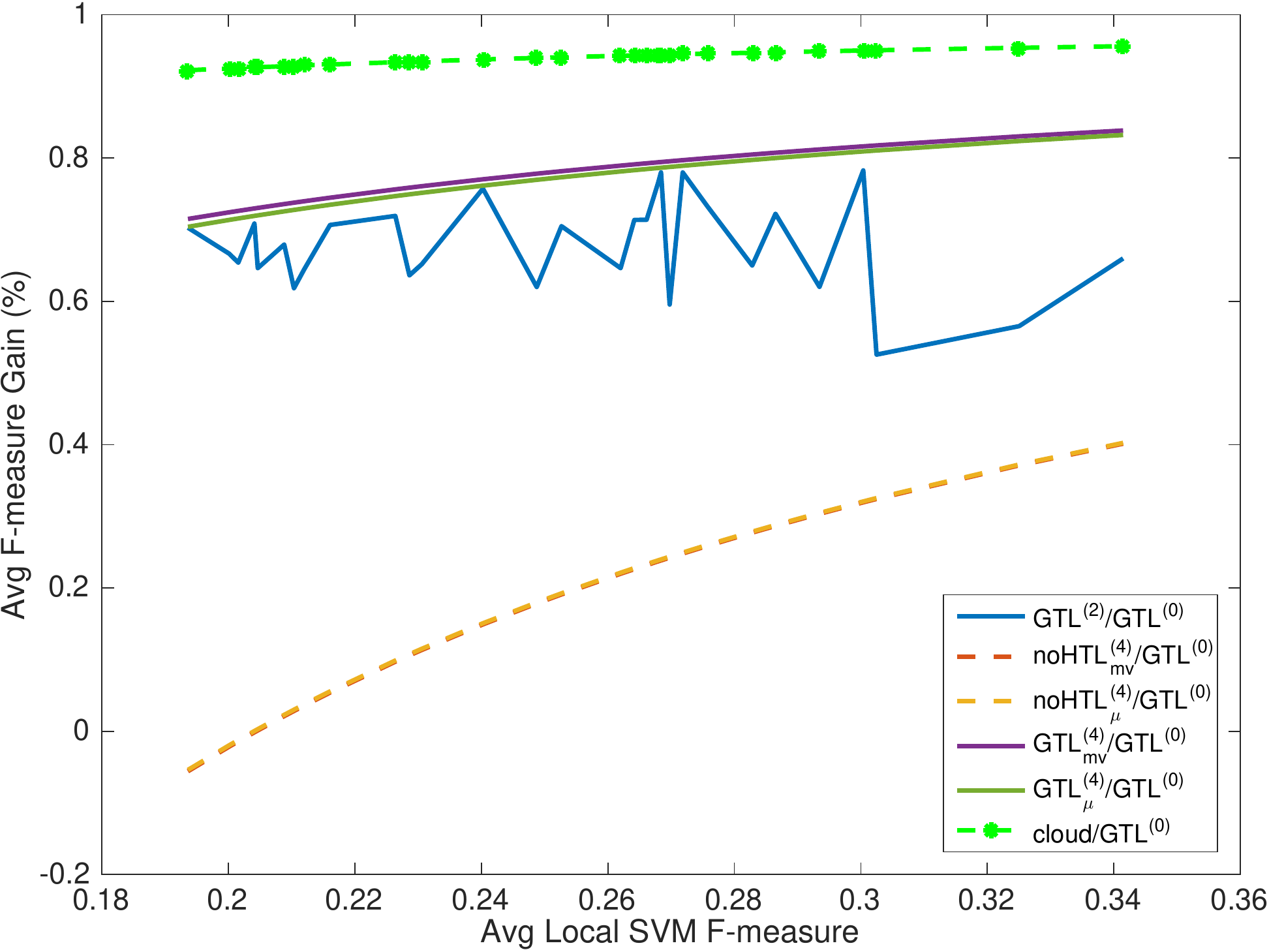}
\label{fig:mnist_unbal_gain}
}
\caption{MNIST with class unbalance. Prediction Performance (a)
and Prediction Performance Gain (b) comparison per single location after each
step  of the considered distributed learning approach for both GTL and noHTL
versions.}
\end{figure}

We now consider a different situation in which data is not
uniformly distributed between users. Here we refer to the situation represented
in Figure~\ref{fig:mnist_36789} where some classes are less represented than
others. In this way we simulate a
 scenario where some types of
information are more difficult to be sensed with respect to others.

Although the type of imbalance is the same as in the HAPT dataset, here
the behaviour of GTL compared to noHTL is different.
Looking at Figure \ref{fig:mnist_unbal_acc} and \ref{fig:mnist_unbal_gain}
we see that after each step of GTL we obtain an improvement in terms of
prediction performance. In this case the transfer of knowledge from one device
to another allows us to obtain in Step 2 ($GTL^{(2)}$) an improvement in prediction
ranging from $50\%$ to $70\%$.  Moreover, after the second phase of model
aggregation in  Step 4 ($GTL^{(4)}$) we obtain a further improvement in
performance in which all devices obtain a final F-measure value of $0.95$.
If we look at the performance of the noHTL solution we notice that the
combination of models is beneficial with respect to the performance of the
local models, but is it not sufficient to overcome poor
representation of some classes at some locations. With respect to the HAPT
dataset, which was also class unbalanced, here GTL outperforms noHTL
also before aggregation (i.e., after Step 2) thanks to transfer of knowledge
across locations. This is due to the fact that poorly represented classes in
this case are even less represented than in HAPT.

\begin{figure}
\centering
\subfloat[]{
    \includegraphics[width=.48\columnwidth]{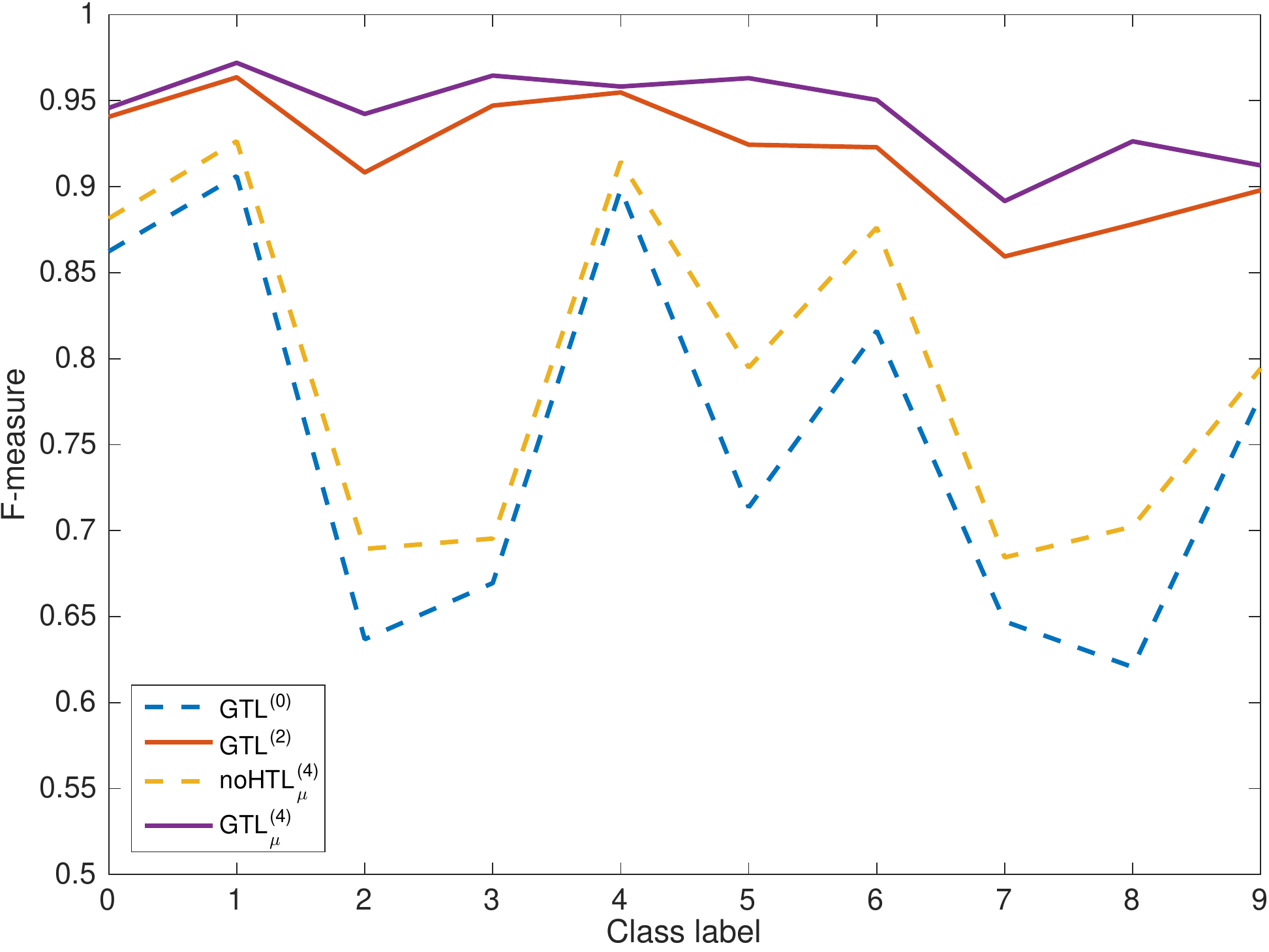}
    \label{fig:mnist_unbal_classAcc}
}
\subfloat[]{
    \includegraphics[width=.48\columnwidth]{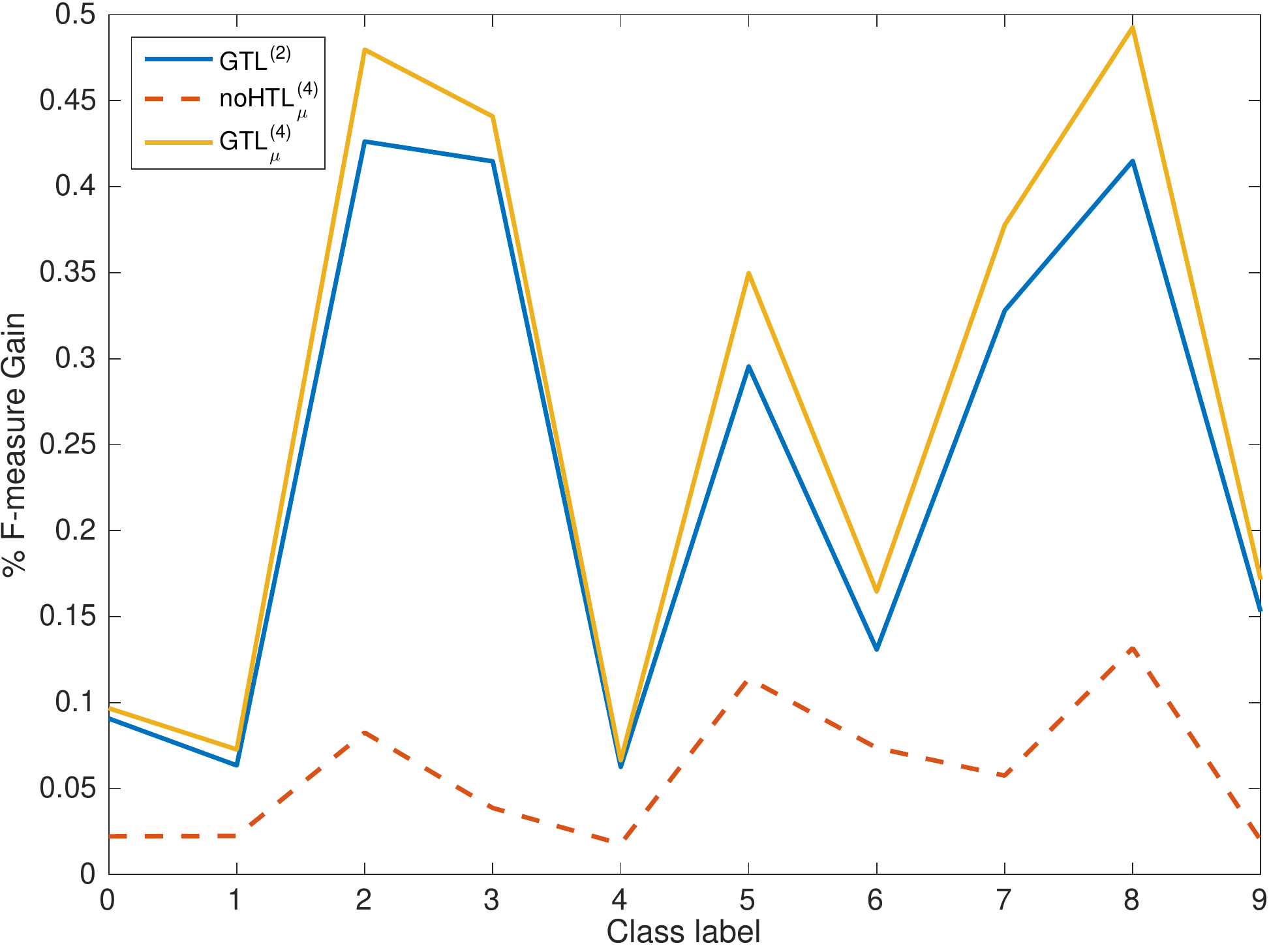}
    \label{fig:mnist_unbal_classAccGain}
}
\caption{MNIST with class unbalance. (a) Prediction Performance
per class before and after the first and the second models' exchange. (b)
Prediction Performance Gain per class after the first and the second model
exchange.}
\end{figure}

Looking at Figures \ref{fig:mnist_unbal_classAcc} and
\ref{fig:mnist_unbal_classAccGain} the benefit of GTL is even more clear. In fact if we look at the prediction
performance per class and the performance gain per class we notice that, thanks
to transfer of knowledge in HTL, each device is able
to learn a model that, on the
one hand, improves the prediction performance of quite well represented classes
and, most importantly,  drastically helps to fill the
performance gap for
those classes the are scarcely represented. The
reason of
this performance improvement is that the knowledge extracted locally from
  poorly represented classes is transferred and aggregated between locations,
and this overcomes the lack of data points at each individual location. In fact,
as shown
in Figure \ref{fig:mnist_unbal_classAccGain} the
performance gain for classes 2,3,5,7,8 is in the range $30\%-50\%$. Regarding
the noHTL solution, in this case simply averaging local models is not sufficient to cope with poorly
represented classes. In fact, we can see some improvement for the classes less
represented but it is not comparable to what obtained with the GTL solution.

\subsection{Case Study IV: MNIST with node unbalance}

\label{sub:one-hot}
\begin{figure}[ht!]
\centering
\subfloat[]{
\includegraphics[width=.48\columnwidth]{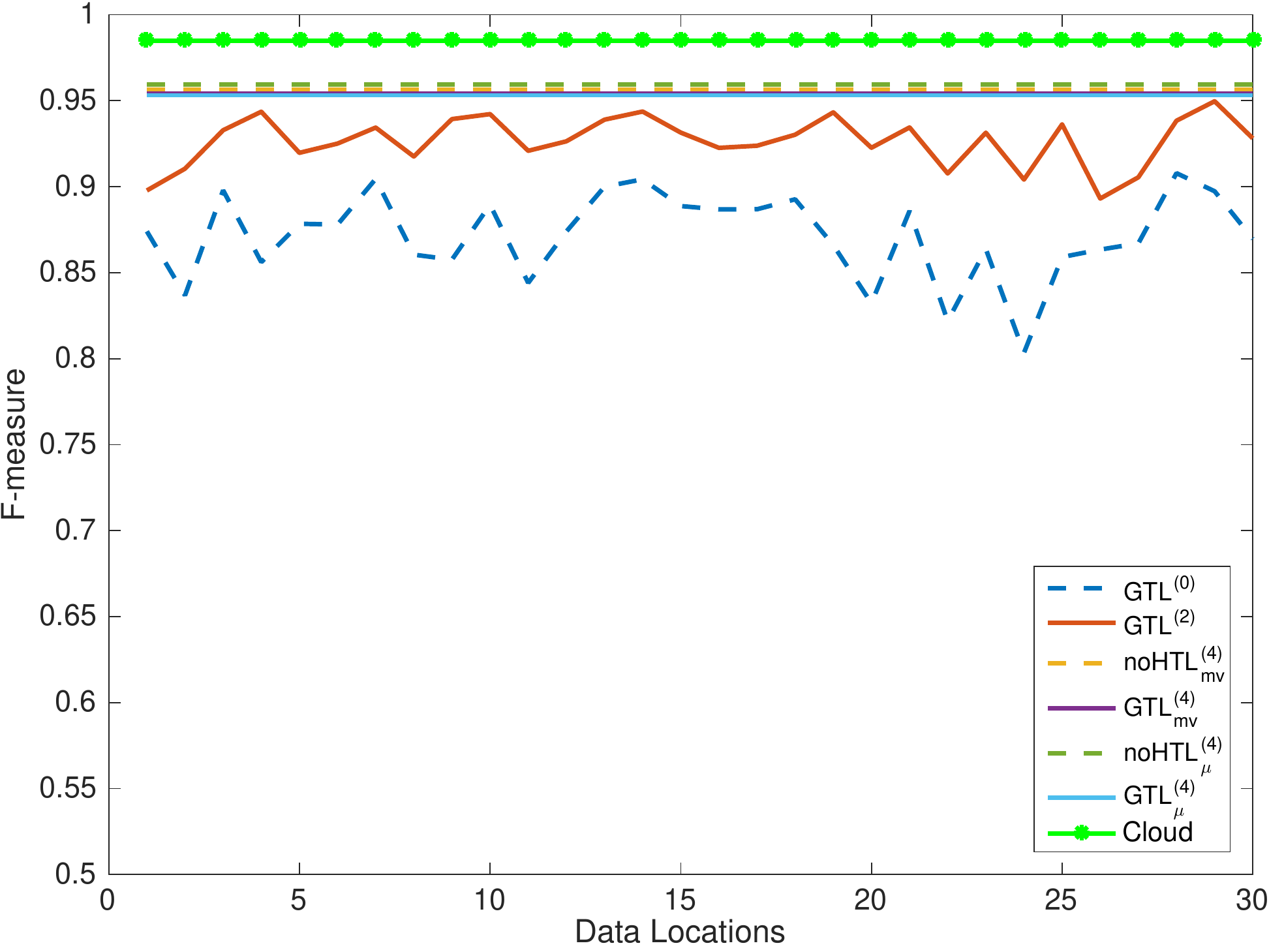}
\label{fig:mnist_oneHot_acc}
}
\subfloat[]{
\includegraphics[width=.48\columnwidth]{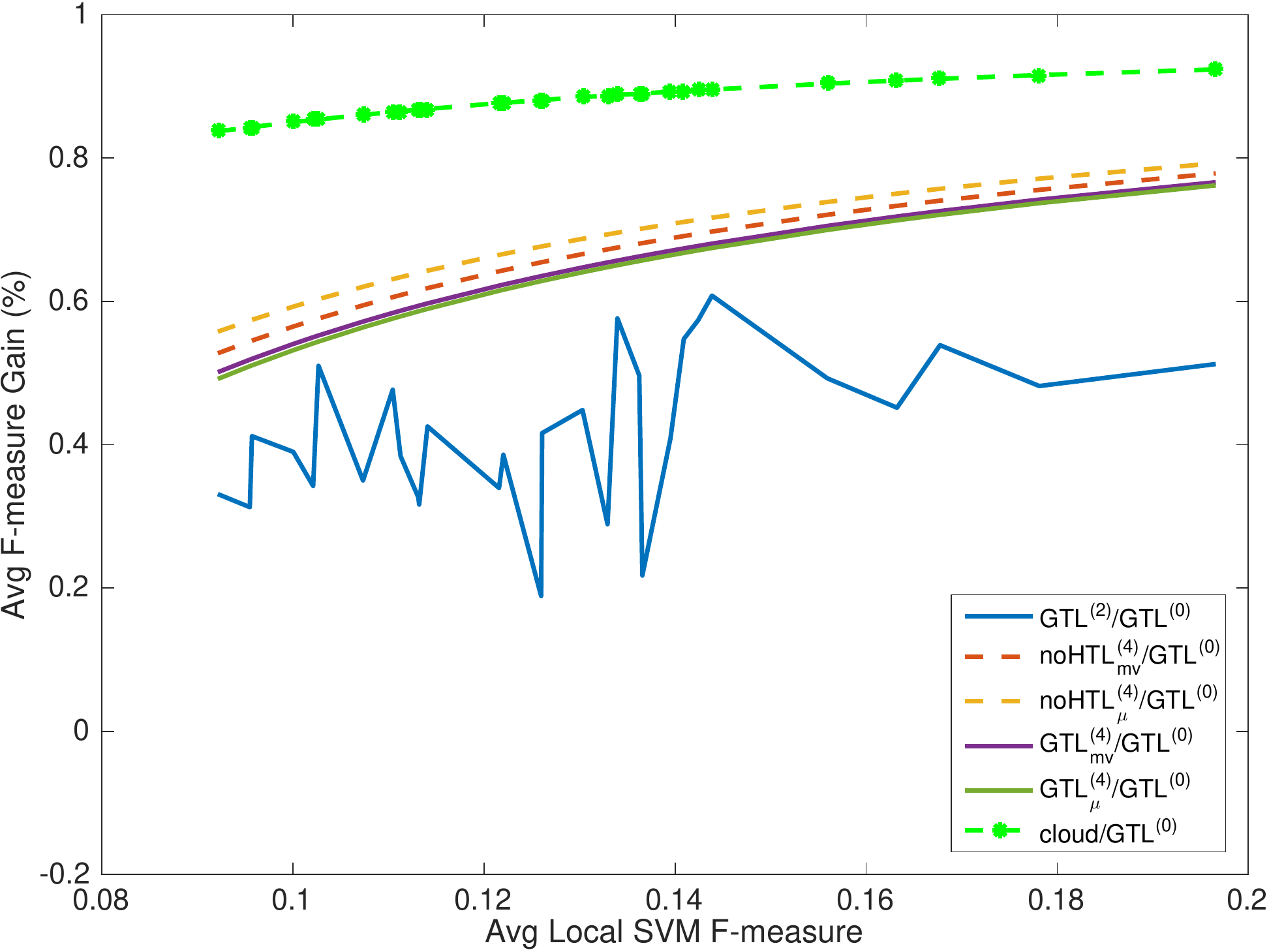}
\label{fig:mnist_oneHot_gain}
}
\caption{MNIST with node unbalance. Prediction performance (a)
and Prediction Performance Gain (b) comparison per single location after each
step  of the proposed distributed learning approach for both HTL and noHTL
versions.}
\end{figure}

Here we want to simulate a limit case where the distribution of classes
collected by user devices is highly skewed. Precisely, in each local dataset
 $70\%$ of data belongs to one class and the remaining $30\%$ is equally
divided between all the other classes. Since we have 30 users and 10 classes,
each class is well represented in $3$ different user devices and only one well
represented class per user is allowed. An example of data distribution for User
1 and User 2 is shown in Figures \ref{fig:onehot1} and \ref{fig:onehot2},
respectively.

\begin{figure}
\centering
\subfloat[]{
    \includegraphics[width=.48\columnwidth]{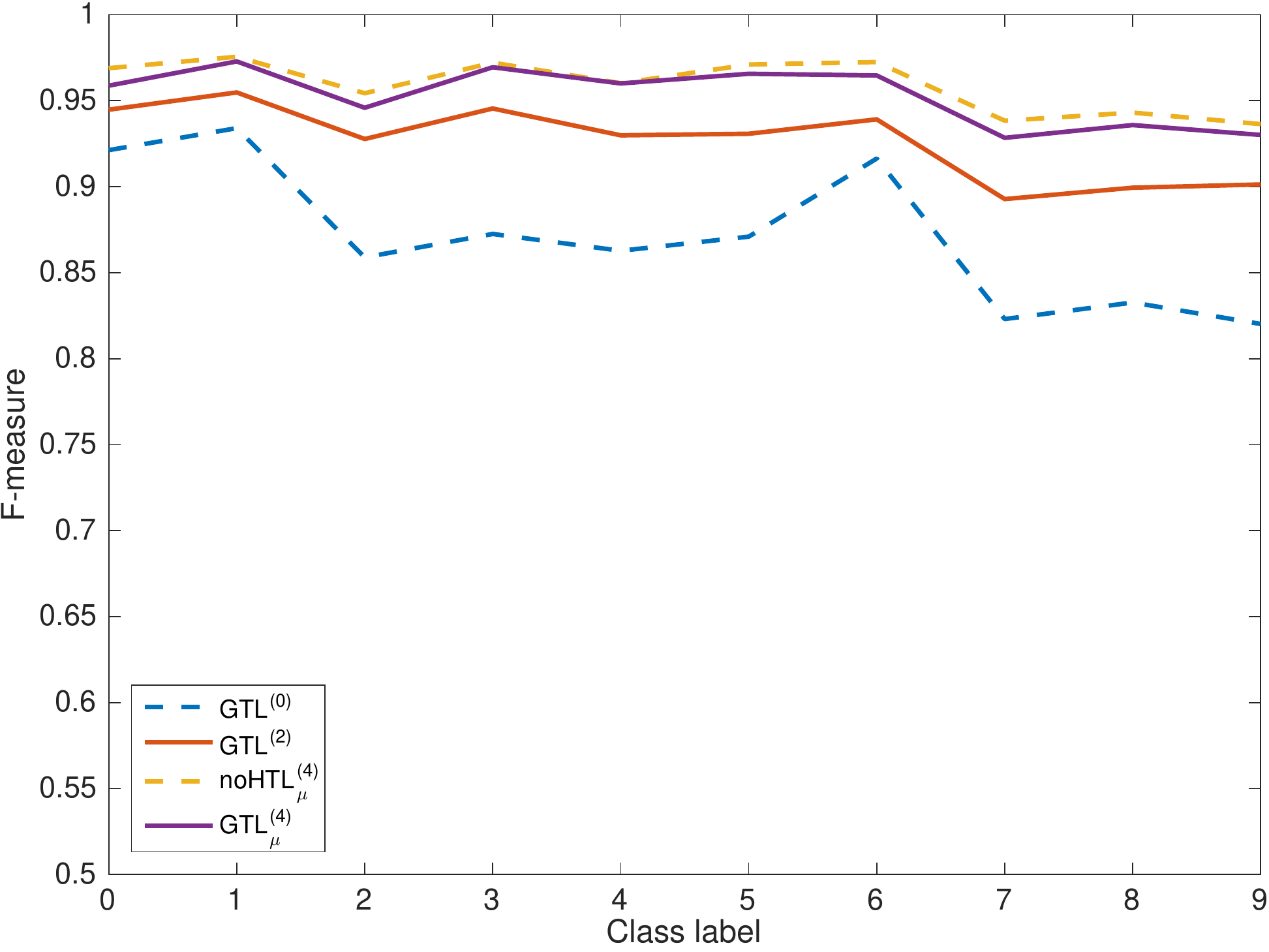}
    \label{fig:mnist_oneHot_classAcc}
}
\subfloat[]{
    \includegraphics[width=.48\columnwidth]{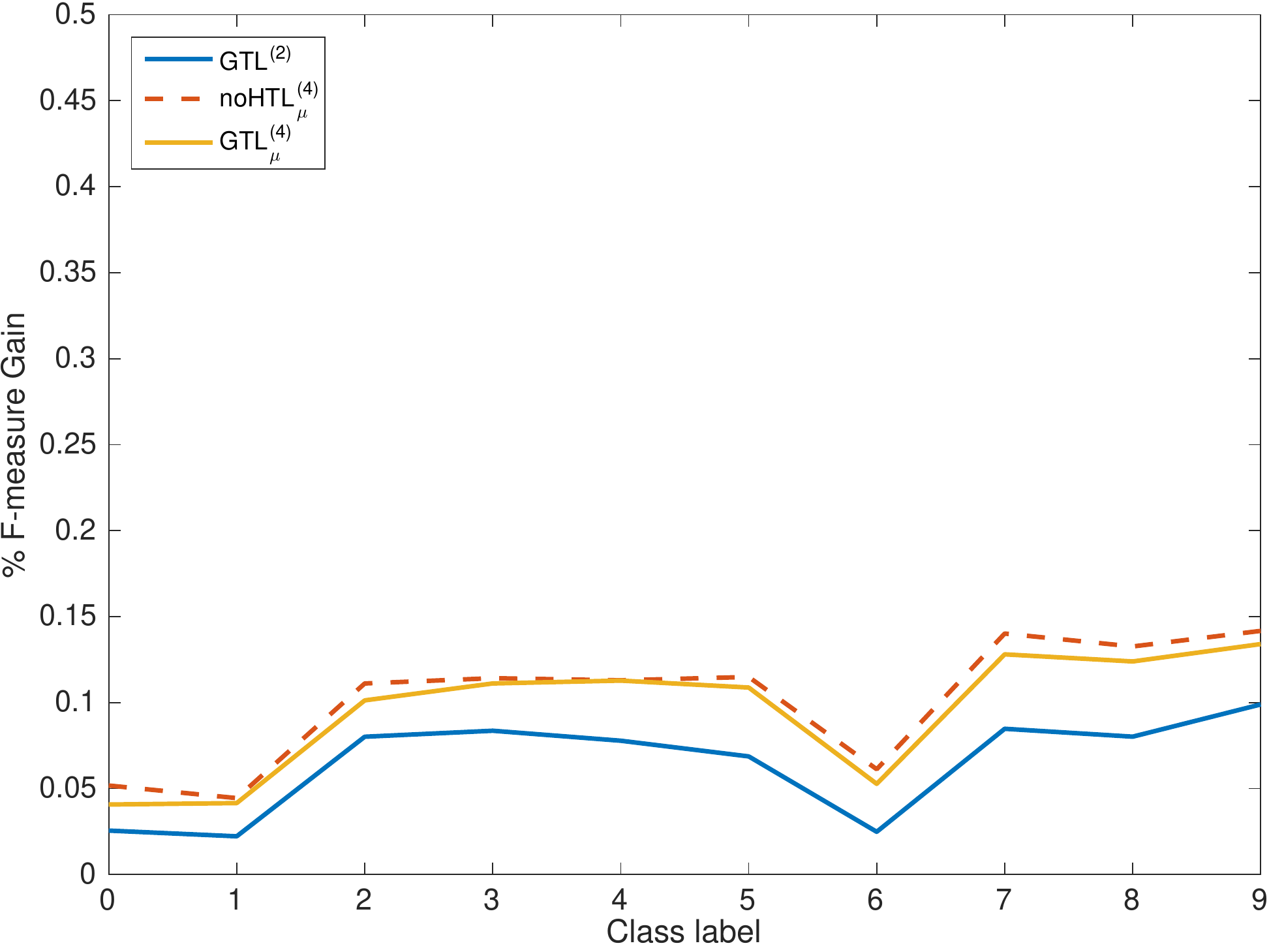}
    \label{fig:mnist_oneHot_classAccGain}
}
\caption{MNIST with node unbalance. (a) Prediction performance
per class before and after the first and the second models' exchange. (b)
Prediction Performance Gain per class after the first and the second model
exchange.}
\end{figure}

Interestingly, as shown in Figure \ref{fig:mnist_oneHot_acc}, in this very
extreme scenario we see that in terms of prediction performance, both GTL
and noHTL approaches
have quite similar performance. In fact they resemble the behaviour obtained in the
balanced dataset, but differently from those results, the performance gain
obtained by both GTL and noHTL approaches are much greater, as depicted in
Figure \ref{fig:mnist_oneHot_gain}.  The very same conclusion can be drawn
also
looking the prediction performance per class and their performance gain
  (Figures~\ref{fig:mnist_oneHot_classAcc} and
  \ref{fig:mnist_oneHot_classAccGain}). Also
  here we see that GTL and noHTL yield almost the same performance, and
  both drastically outperform the local models. This tells that, in case of node
unbalance, distributed learning is able to ``re-balance'' representativeness of
classes across nodes, ultimately yielding the same performance that also local
model would obtain in case of completely balanced datasets.

\section{Robustness to malicious behaviour}
\label{sec:robust}
In order to further investigate the suitability of the proposed distributed learning solutions  we evaluate its robustness in presence of  ``malicious'' devices. Precisely, we consider a scenario in which $s'$ out of $s$ devices involved in the learning process are ``malicious'' and their purpose is to affect negatively the accuracy of the model learnt during the distributed learning process. In order to simulate such scenario we assume that during  Step 1 of the GTL  and noHTL procedures, the partial models sent by the $s'$ devices are corrupted. 
In this analysis a model can be partially or totally corrupted. In order to corrupt a model, we randomly select a subset of its parameters and we replace them with random numbers generated according to a normal distribution with zero mean and unitary standard deviation.  In order to identify the model's parameters to be replaced (i.e., corrupted), for each parameter we perform a random trial according to a Bernoulli distribution with parameter $p$. Through the parameter $p$  we control (on average) the percentage of corrupted model's parameters, e.g., if $p=0.25$, 25$\%$ of the nodes (on average) will produce corrupted models.
We are interested in quantifying the prediction performance degradation induced by the injection of corrupted model in the distributed learning process.

In our simulation we identify two separate scenarios, called respectively Malicious1 and Malicious2. In the Malicious1 scenario,  the malicious devices send to all the other nodes a fully corrupted partial model, i.e., all the model's parameters are random values. In order to evaluate the robustness of both GTL and noHTL in presence of such type of attacks, we varied the percentage of malicious devices, i.e., $25\%,50\%,75\%$.  Results are presented in Table \ref{tab:noisesrc-mnist} and \ref{tab:noisesrc-hapt}.
We notice a drastic difference in performance between GTL and noHTL for both the datasets considered.  GTL proves to be more robust to the presence of malicious nodes. The main reason  is that with GTL, devices are able to i) identify and select the most informative sources (i.e. the partial models with a real informative content) and, most importantly, ii)  exclude automatically  partial models that do not contain useful knowledge. Clearly, when the number of malicious devices (and models) increases too much, the performance of GTL slightly decreases but it still remains very good. Conversely, as we may expect, the performance of noHTL is stronlgy affected by the precence of ``malicious'' partial models. In fact, the noHTL procedure averages the partial models in order to come up with a final model, thus when the number of malicious models outnumbers the ``good'' ones, its performance degrades accordingly.  
\begin{table}[ht!]
\centering
\caption{MNIST balanced. F-Measure of GTL and noHTL in the Malicious1 scenario.\label{tab:noisesrc-mnist}}
\begin{tabular}{|c|c|c|}
\hline
$\%$ Malicious devices & noHTL$_{\mu}$ & GTL$_{\mu}$\\
\hline
$25\%$ & $0.94$ & $0.970$ \\
$50\%$ & $0.72$ & $0.972$ \\
$75\%$ & $0.40$ & $0.971$ \\
\hline
\end{tabular}
\end{table}
\begin{table}[ht!]
\centering
\caption{HAPT balanced. F-Measure of GTL and noHTL in the Malicious1 scenario.\label{tab:noisesrc-hapt}}
\begin{tabular}{|c|c|c|}
\hline
$\%$ Malicious devices & noHTL$_{\mu}$ & GTL$_{\mu}$\\
\hline
$25\%$ & $0.83$ & $0.970$ \\
$50\%$ & $0.56$ & $0.95$ \\
$75\%$ & $0.24$ & $0.95$ \\
\hline
\end{tabular}
\end{table}

In the Malicious2 scenario we consider the case in which all the devices are malicious but the partial models they exchange with each other during Step 1 are only partially corrupted. Specifically we evaluate both GTL and noHTL varying the corruption level of exchanged models. In our simulations, the percentage of corrupted parameters at each node is $25\%,50\%,75\%$. Note that this case also represents non-malicious scenarios, where for any other reasons the partial models parameters are inaccurate or corrupted. Results are reported in Tables \ref{tab:noisemdl-mnist} and \ref{tab:noisemdl-hapt}. Similarly to the Malicious1 scenario, we see that GTL is able to select which parameters are informative and which are not, thus preserving an overall good prediction performance. Regarding noHTL we see also in this case that the presence of corrupted parameters strongly affects its prediction performance. 

\begin{table}[hb!]
\centering
\caption{MNIST balanced. F-Measure of GTL and noHTL in the Malicious2 scenario.\label{tab:noisemdl-mnist}}
\begin{tabular}{|c|c|c|}
\hline
$\%$ of corrupted params. & noHTL$_{\mu}$ & GTL$_{\mu}$\\
\hline
$25\%$ & $0.90$ & $0.97$ \\
$50\%$ & $0.77$ & $0.97$ \\
$75\%$ & $0.43$ & $0.96$ \\
\hline
\end{tabular}
\end{table}
\begin{table}[hb!]
\centering
\caption{HAPT balanced. F-Measure of GTL and noHTL in the Malicious2 scenario.\label{tab:noisemdl-hapt}}
\begin{tabular}{|c|c|c|}
\hline
$\%$ of corrupted params. & noHTL$_{\mu}$ & GTL$_{\mu}$\\
\hline
$25\%$ & $0.78$ & $0.960$ \\
$50\%$ & $0.43$ & $0.96$ \\
$75\%$ & $0.29$ & $0.96$ \\
\hline
\end{tabular}
\end{table}

\section{Network overhead performance}
\label{sec:noh}
We hereafter focus on the analysis of the network traffic generated by
the different learning schemes considered in the paper.
With network overhead (OH) we refer to the amount of information that must
be sent between all the data locations in order to accomplish the learning task.
We recall that in the GTL solution, local information is sent over the network
twice, i.e. to send the GTL$^{(0)}$ models and to send the GTL$^{(2)}$ models
after Step 2. Therefore, the amount of information sent in GTL is given by the
size of the coefficients of the model in the two steps. This is $\omega^{(0)}$
and $\omega^{(1)}$, respectively. Specifically, $\omega^{(0)}$ is a vector of $d$ coefficients
where $d$ is the dimensionality of the data points ${\bf x}\in {\bf X}$. On the
other hand, it holds that $\omega^{(1)} \in \mathbb{R}^{d+s-1}$, where $s$ is the
number of data locations. Table~\ref{tab:notation} reports the relevant
notation.
\begin{table}[H]
    \footnotesize
\centering
\caption{Network overhead notation\label{tab:notation}}
\begin{tabular}{|l|c|}
    \hline
    $s$ & n. of data locations \\
    $k$ & n. of classes \\
    $N$ & n. of data points in the entire dataset\\
    ${\bf \omega}^{(0)}\in \mathbb{R}^d$ & coefficients of the SVM model\\
    ${\bf \omega}^{(1)}\in \mathbb{R}^{d+s-1}$ & coefficients of the
    GTL model\\
    $d^{(0)}$  & n. of non-null coefficients in $\omega^{(0)}$\\
    $d^{(1)}$ & n. of non-null coefficients in $\omega^{(1)}$\\
    \hline
\end{tabular}
\end{table}

 The network overhead generated by  GTL is computed as follows:
\begin{equation} 
    \label{eq:genNOH}
    OH^{GTL} = OH^{(0)} + OH^{(1)}
\end{equation}
where
\begin{eqnarray}
    OH^{(0)} &= s(s-1)d^{(0)}k\label{eq:ohsvm}\\
    OH^{(1)} &=  s(s-1)d^{(1)}k\label{eq:ohgtl}
\end{eqnarray}

With respect to $OH^{(0)}$, the expression comes from the fact that after
step 1 each location has to send to the other $s-1$ locations a classifier for
each class. As each classifier consists in $d^{(0)}$ \emph{non-null} coefficients, and we have $k$
classes, the formula is straightforward. With respect to $OH^{(1)}$, in step 3
each location has to send again a classifier per class, but this time the
classifier consists in $d^{(1)}$ \emph{non-null} coefficients.


Table \ref{tab:noh} reports the network overhead for each models' exchange
phase of the GTL procedure. As we can see, the most expensive phase is the
exchange of the local
models. The second one, instead, has a negligible cost if compared to the first
one. The main reason  is the fact that the number of non-null
coefficients after the first GTL step (i.e., Step 2 in the algorithm) is
typically much smaller than the number of non-null coefficients of the
local models. This is a general property of GreedyTL, that also works as a feature
selection algorithm \cite{OrabonaGreedyTL}. Remember that in our
dataset data points are not raw data, but are features extracted locally
from the raw data. In most IoT applications, instead, knowledge extraction
would be done starting from the raw data, which will significantly impact the overhead of the Cloud solution. Specifically, the table also presents the
  overhead of the Cloud solution, in both cases where features ($OH^{cl}$) or
  raw data ($OH^{raw}$) are
collected. These results clearly indicate that using GTL as
opposed to a centralised cloud solution achieves a drastic reduction in
terms of network overhead, which amounts to 52\% and 83\% in the case of the
HAPT and MNIST datasets, respectively.
Note that if we consider the raw data instead of pre-processed data (as it
would be common for many applications) the reduction in overhead would jump
to 77\% and 83\% for HAPT and MNIST datasets, respectively. This is a extremely
good result, if one considers the obtained prediction performance.

\begin{table}[ht]
    \footnotesize
    \centering
    \caption{Empirical network overhead for the GTL solution}
    \label{tab:noh}
    \begin{tabular}{|l|l|l|l|l|l|l|l|}
        \hline
        & $OH^{(0)}$ & $OH^{(1)}$ & $OH^{tot}$ & $OH^{cl.}$ & $OH^{raw}$& Gain & Gain$_{raw}$\\
        \hline
        HAPT & $20$MB & $3$MB & $23$MB & $48$MB & $103$MB &  $52\%$ &$77\%$\\
        MNIST & $21$MB & $4$MB & $25$MB & $148$MB & $358$MB &  $83\%$ &$93\%$\\
        \hline
    \end{tabular}
\end{table}

Let us now consider the network overhead performance of the noHTL solution. 
The network overhead triggered by the noHTL distributed learning procedure
depends on the models' aggregation method. In fact, if we want to compute the
mean model (Consensus-based aggregation), it is sufficient to
collect all models  $\omega^{(0)}$ into one device, compute the mean model and
send it back to all the other devices. Conversely, if we consider to use a
majority voting approach, all the  devices must have all the $\omega^{(0)}$ models.
The network overhead generated by noHTL is computed as follows:
\begin{eqnarray}
    OH_{\mu}^{noHTL} &=& 2k(s-1)(d^{(0)})\label{eq:oh_munohtl}\\
    OH_{mv}^{noHTL} &=& ks(s-1)d^{(0)}\label{eq:oh_mvnohtl}
\end{eqnarray}
where $\bar{d}^{(0)}$ is the size of the mean model sent back to all devices.
It is clear that Consensus-based aggregation is more efficient, as it
cuts traffic by a factor equal to the number of locations, $s$.

Table~\ref{tab:noh_nohtl} shows the performance of both versions of
noHTL. It is confirmed that using Consesus-based aggregation is way more
efficient (one order of magnitude) than Majority voting. As, in terms of
prediction performance, there is no significant different between the two,
typically the former would be preferred. Also note that,  as expected, the
overhead of noHTL with Majority voting is comparable with that of GTL.

Therefore, we can conclude that (i) both GTL and noHTL provide a drastic
reduction of overhead with respect to Cloud-based solution, also when data are
pre-processed locally to extract features; (ii) GTL generates more traffic in
the network with respect to noHTL with Consesus-based aggregation, but there are
cases where it can provide a very significant gain in terms of prediction
performance. The choice between GTL and noHTL depends thus on the specific
requirements in terms of predictivity vs. network traffic trad-off. We come back
on this point in Section~\ref{sec:sensitivity}, where we show how to
significantly reduce the
GTL overhead without affecting its prediction performance.
 \begin{table}[ht]
     \footnotesize
     \centering
     \caption{Empirical network overhead for noHTL solution.}
     \label{tab:noh_nohtl}
     \begin{tabular}{|l|l|l|l|l|l|l|l|l|}
         \hline
         & $OH_{\mu}^{noHTL}$ & $OH_{mv}^{noHTL}$ & $OH^{cl.}$ & $OH^{raw}$&
	 Gain$_{\mu}$ & Gain$_{\mu}^{raw}$& Gain$_{mv}$ & Gain$_{mv}^{raw}$\\
         \hline
         HAPT & $2$MB & $20$MB & $48$MB & $103$MB &  $96\%$ &$98\%$ & $58\%$ &
	 $81\%$\\
         MNIST & $1.45$MB& $21$MB &$148$MB & $358$MB &  $99\%$ &$99.5\%$ &
	 $86\%$ & $94\%$\\
         \hline
     \end{tabular}
 \end{table}

\subsection{Bound on GTL network overhead}
In order to deeply understand the behaviour of the GTL solution from the point of
view of the network overhead, we derived, under some reasonable assumption, an
upper bound of the network overhead. Our only assumption is that the model
learnt by GreedyTL at Step 2 has less non-zero elements than the base model.
Formally, provided that  $$d^{(1)} \leq d^{(0)}$$ 
and assuming, as it is   typically the case, that the number of non-null coefficients of the   learners ($d^{(0)}$ and $d^{(1)}$) is much higher than the number of locations $s$, 
the network overhead is upper bounded as follows:

\begin{equation}
  OH^{tot} \leq OH^{up} \triangleq 2ks^2d^{(0)}\,\,.
\label{eq:bound}
\end{equation}

To obtain Equation~\ref{eq:bound} we note that in general we can
approximate $OH^{(0)}$ as $s^2 d^{(0)} k$. Under the assumption $d^{(1)} \gg s$,
$OH^{(1)}$ can be approximated as $s^2 d^{(1)} k$, and because $d^{(1)} \leq
d^{(0)}$, $OH^{(1)} \leq OH^{(0)}$ holds true.  Therefore, the following also
holds:

    \begin{equation}
\begin{split}
    OH^{(0)} + OH^{(1)} &\leq 2*OH^{(0)} \\
    &\leq 2k(s(s-1)d^{(0)}) \\
    &\leq 2ks^2d^{(0)} - 2k s d^{(0)} \\
     &< 2ks^2d^{(0)}
\end{split}
\end{equation}

Note that the bound in Equation~\ref{eq:bound} can be quite pessimistic. In many
cases (as with our dataset), actually $d^{(1)} \ll d^{(0)}$ holds, and therefore
$OH^{tot}$ can be well approximated with $OH^{(0)} \simeq s^2 k d^{(0)}$. However, in the following
we use the more exact (but pessimistic) bound given by Equation~\ref{eq:bound},
thus presenting worst-case results for the network overhead with GTL.

\begin{figure}[ht]
    \centering
    \subfloat[]{
    \includegraphics[width=.30\textwidth]{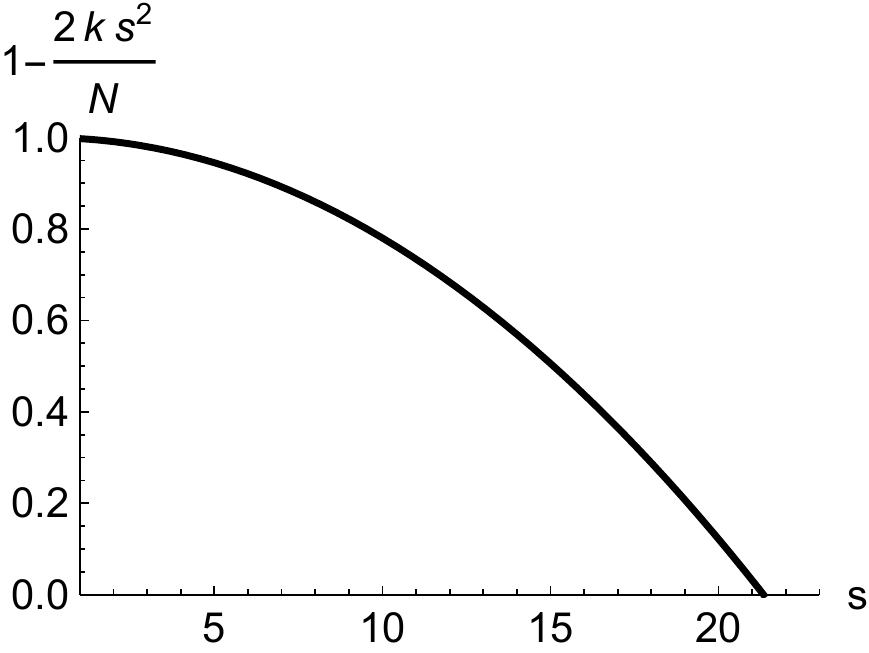}
    \label{fig:noh-s}
}
    \subfloat[]{
    \includegraphics[width=.30\textwidth]{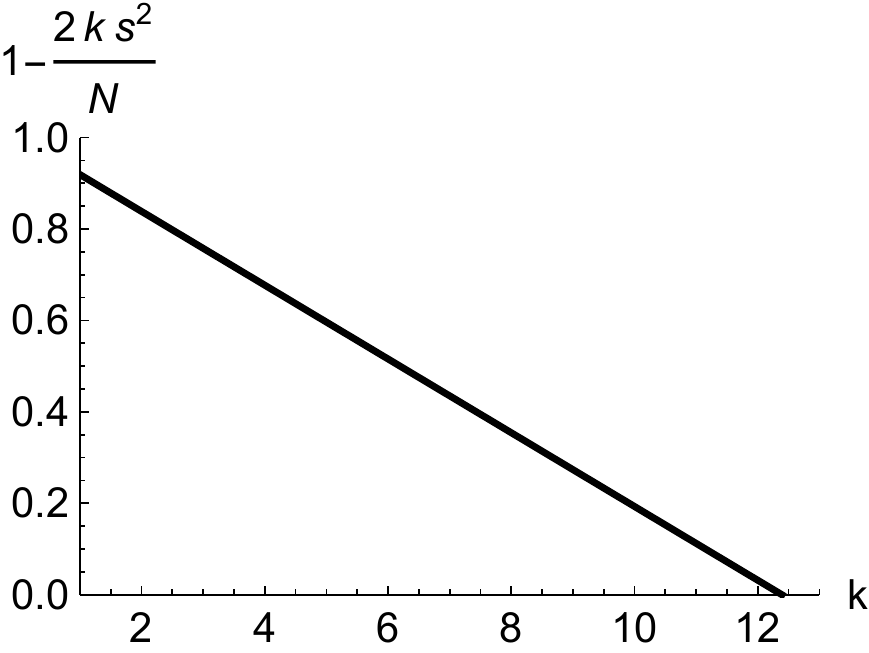}
    \label{fig:noh-k}
}
    \subfloat[]{
    \includegraphics[width=.30\textwidth]{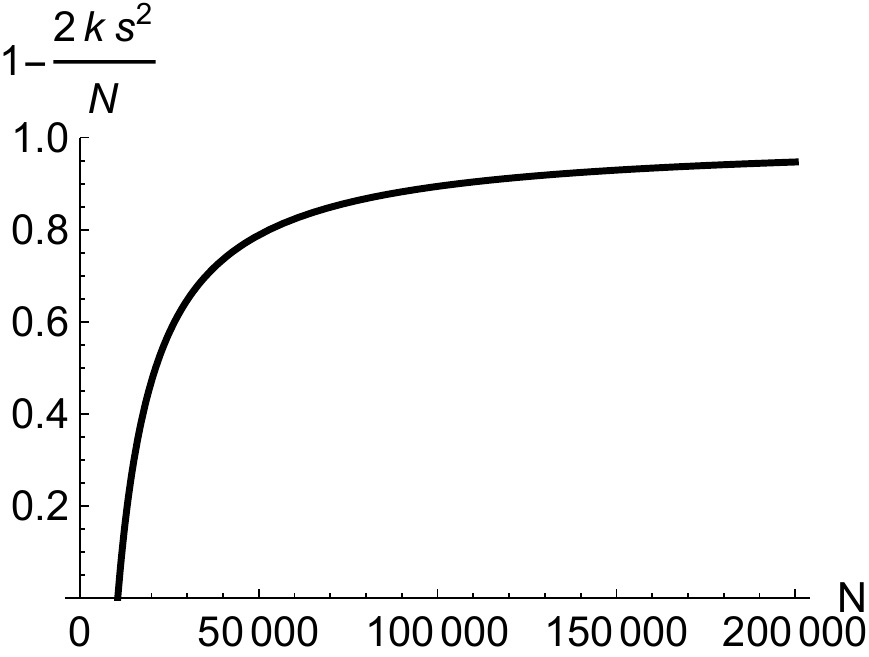}
    \label{fig:noh-N}
}
\caption{Parameters' sensitivity analysis of the network gain with GTL.\label{fig:nohsensan}}
\end{figure}
We can exploit Equation \ref{eq:bound} to derive a lower bound on the
network gain. Specifically, considering that in the case of a centralised cloud
based system the overhead is $OH^{cloud}=Nd^{(c)}$ where $d^{(c)}$ is the
dimensionality of each data point, we can define a lower bound
on the gain as follows
\begin{equation}
  \underline{G} = 1-\frac{OH^{up}}{OH^{cloud}} = 1-\frac{2 k s^2 d^{(0)}}{N
    d^{(c)}} \,\,.
\label{eq:nohgain}
\end{equation}
We use $\underline{G}$ to perform a sensitivity analysis
of the network gain with GTL. Note that in the following we assume, as it
is the case of our datasets, that the dimensionality of the original data points
is approximately the same of the dimensionality of the $GTL^{(0)}$ coefficients
(note that in our dataset the dimensionality of raw data would be even larger,
thus our analysis provides an even lower bound on the possible gain of GTL), and
thus $\underline{G} \simeq 1- \frac{2ks^2}{N}$holds. Precisely in order to
study the impact of each parameter, for each graph in the following we
keep two parameters fixed (we use the values obtained by simulations) and we
vary the remaining one.
From Fig. \ref{fig:noh-k} we see that the  gain is linear with the number of
classes $k$. This is quite obvious because keeping fixed the size of the dataset and the number of data locations, more classes means more binary models to be sent over the network.
Figure \ref{fig:noh-N} is more interesting, as it shows that the more the
size of the dataset ($N$) increases, the more it becomes advantageous to use
GTL, as the relative cost of exchanging models instead of data becomes
negligible. This is also the case we have observed in simulation,
  reported in Table~\ref{tab:noh}: the MNIST dataset is larger than HAPT, and
this results in a higher gain in using GTL. Also note that the curve is concave, and thus also for relatively
small data sizes (with respect to the number of locations and dimensionality of our
dataset) the cost of GTL becomes already almost negligible.

Conversely, looking at Figure \ref{fig:noh-s}, we see that there is a
limit on the
number of data  locations, above which the exchange of model
coefficients becomes too
expensive. Let us define as $\mu_D$ the average number of data points per
location, such that $N=s \mu_D$. The gain can then be written as follows:
\begin{equation}
  \underline{G} = 1-\frac{2 k s^2 d^{(0)}}{N d^{(c)}} \sim 1 - \frac{2k s^2}{s \mu_D}  = 1 - \frac{2k s}{\mu_D}
\end{equation}
This form of $\underline{G}$ has an intuitive explanation. For each
location, $ks$ is the network overhead of using GTL, as it is the total size of
models' parameters to be sent. On the other hand, $\mu_D$ is the per-site cost
of a cloud-based solution. As soon as the number of locations increases above
$\frac{\mu_D}{2k}$ using GTL becomes not advantageous anymore.

\section{Tuning the prediction vs. network overhead
trade-off}
\label{sec:sensitivity}
\begin{figure}
\centering
\subfloat[MNIST balanced]{
\includegraphics[width=.4\textwidth]{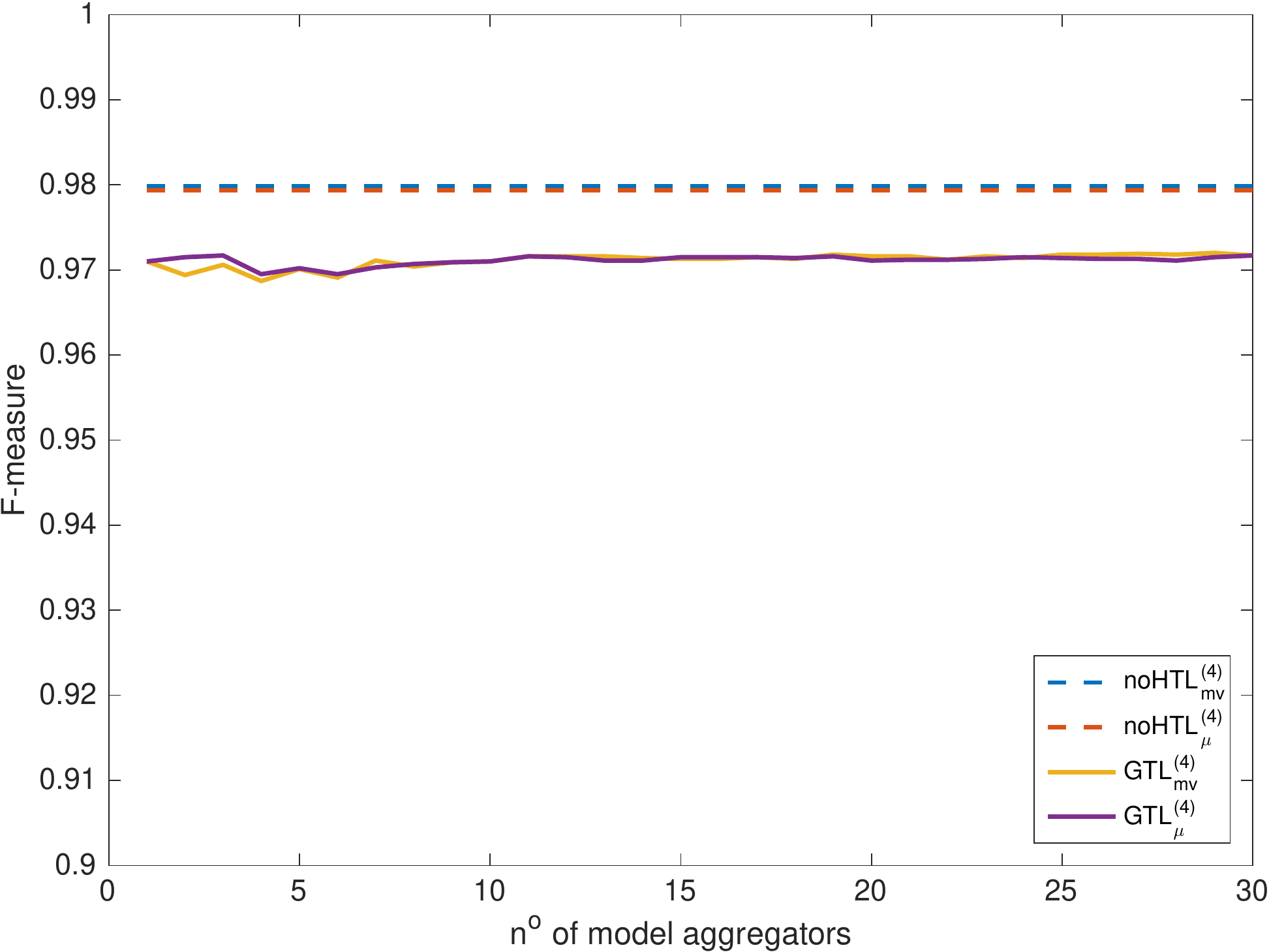}
\label{fig:mnist_incloss_bal}
}
\subfloat[MNIST class unbalance]{
\includegraphics[width=.4\textwidth]{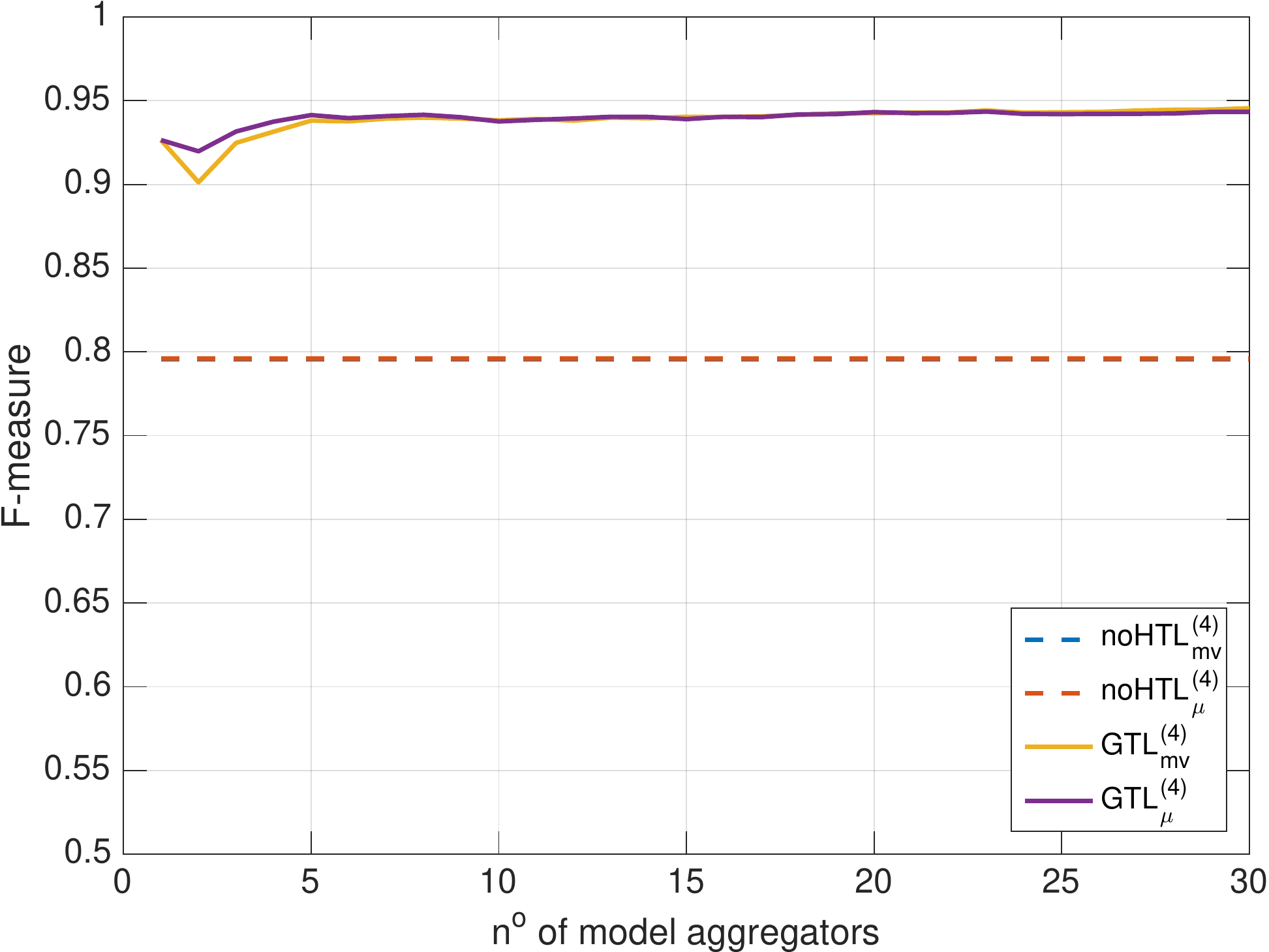}
\label{fig:mnist_incloss_ubal}
}\\
\subfloat[HAPT]{
\includegraphics[width=.4\textwidth]{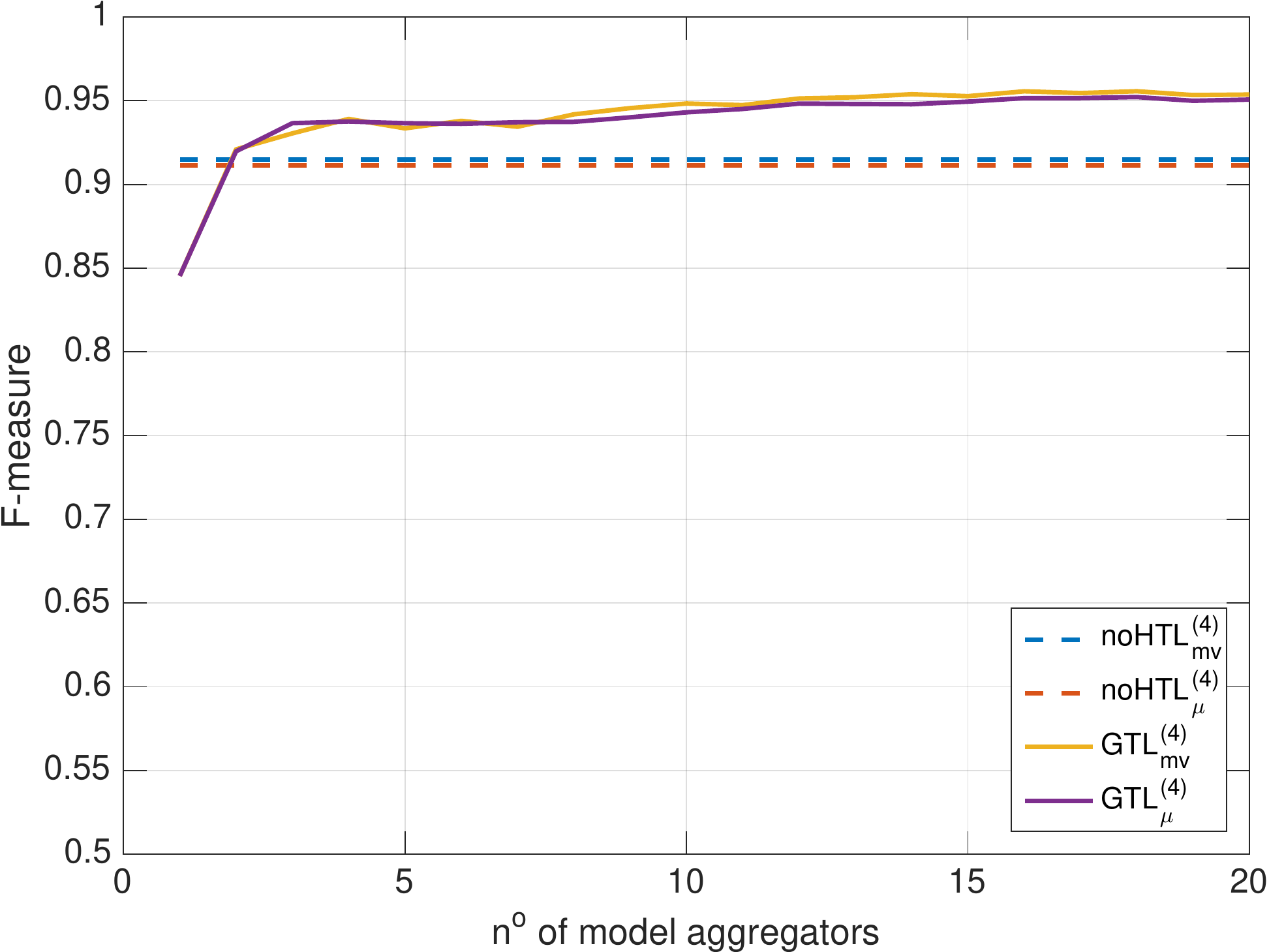}
\label{fig:hapt_incloss}
}
\subfloat[MNIST node unbalance]{
\includegraphics[width=.4\textwidth]{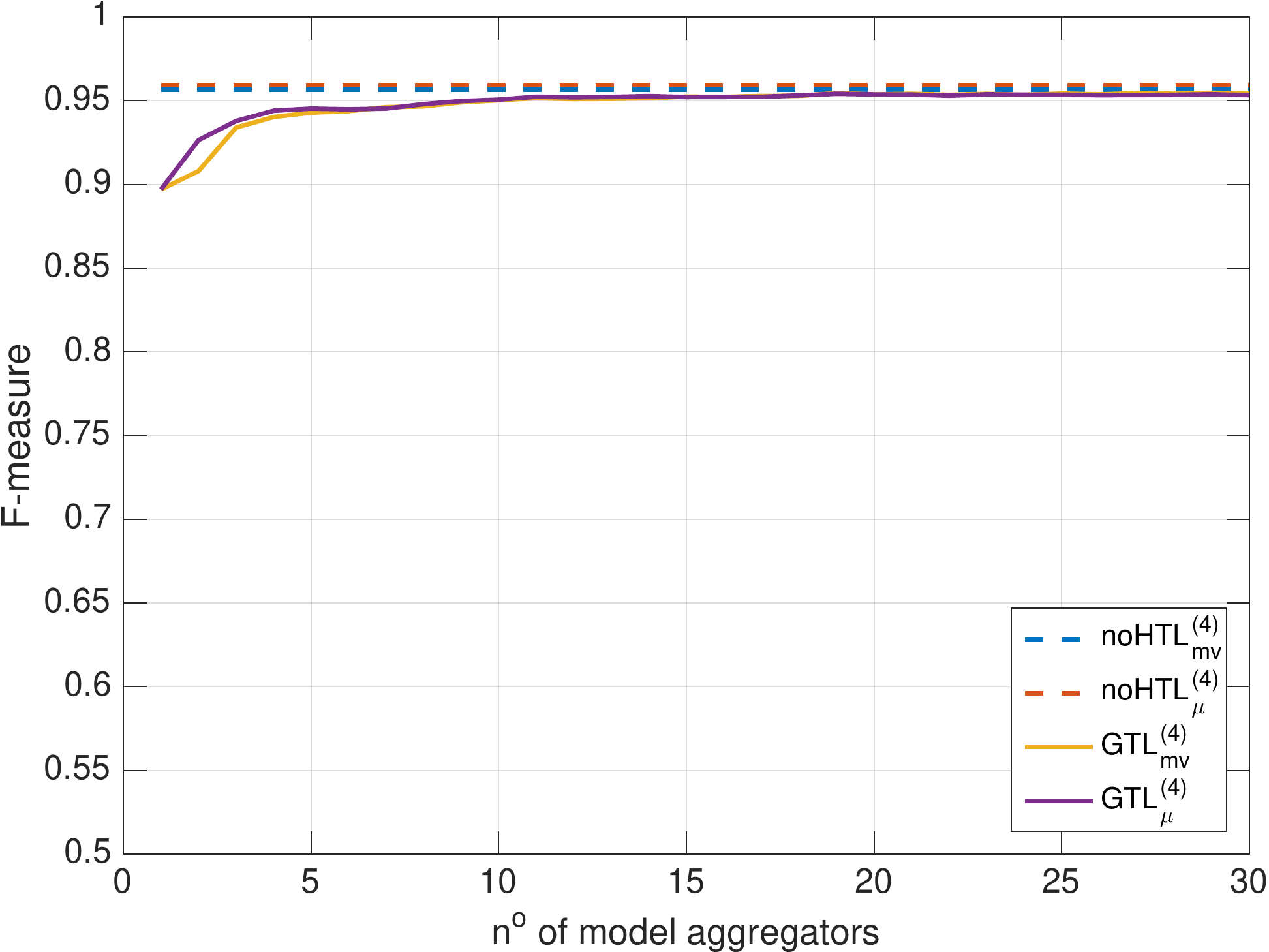}
\label{fig:mnist_incloss_onehot}
}
\caption{Prediction performance increasing the number of
  aggregators involved in the distributed learning process for
  both GTL and noHTL. \label{fig:incloss}}
\end{figure}

From the results presented in Section~\ref{sec:noh} it becomes clear
that both GTL and noHTL have pros and cons. In general
noHTL$_{\mu}$ is the best choice in terms of network overhead, however it is not
always the best solution in terms of prediction performance. As shown before, it is highly dependent on the distribution of classes between different locations. 
Moreover, from the network overhead point of view, the GTL and noHTL$_\mu$
approaches represent the upper and lower limits of the traffic that we can
generate with such type of distributed learning system. One interesting question
 is if there exist an intermediate solution in terms of models
exchange that limits the traffic generated but leaves
 the prediction performance unaffected.

In order to shed some light on this point, for all the scenarios presented
before we performed a sensitivity analysis adopting the following approach.
Regarding GTL, instead of performing the models' synchronisation in parallel for
all devices, we do it for one device at a time in an incremental fashion.
We pick a certain number of nodes as \emph{model aggregators}. This
number can vary between 1 and the total number of nodes. All nodes compute their
local models as per original Step 0. We then send the local models only to the
aggregators, which perform Step 2 of the algorithm, thus computing an aggregate
model based on the ones received, using the local dataset for training. Finally,
only the aggregators perform Steps 3 and 4, i.e., they exchange and aggregate the
models they have computed in Step 2. The final model is sent back to all nodes.
In terms of overhead, when the number of aggregators is equal to 1, this is
equivalent to noHTL with Consensus-based aggregation. When we use all nodes as
aggregators, this is equivalent to the original GTL algorithm.

In Figure~\ref{fig:incloss} we show to what extent the performance of GTL
depends on the number of aggregators involved in the procedure, for each different
experimental setup we presented before. 
For comparison, we also report the performance of noHTL, which clearly
  does not depend on the number of GTL aggregators. More precisely, it would be
  possible to define a scheme for noHTL$_{mv}$ by exchanging only a limited
  number of models. We have decided to show here the performance of the original noHTL$_{mv}$ scheme,
  as a worst-case comparison for GTL. For the case where the dataset is balanced
  (Figure \ref{fig:incloss}(a)), there is no significant gain in using
  additional model aggregators. More interesting is the case of unbalanced
  datasets. In all cases GTL benefits from an additional number of aggregators.
  However, it is very interesting to note that only a small number of
  aggregators is sufficient to already obtain the prediction performance of GTL
  with complete exchange of models across all locations. Therefore, it is
  possible to appropriately tune GTL to outperform noHTL in terms of prediction
  performance, while approximating very closely the performance of noHTL$_\mu$
in terms of network overhead.

\section{Dynamic scenario}
\label{sec:dyn}
In order to complete the analysis of the distributed learning approaches considered in this paper, in this section we evaluate  their applicability to a more dynamic scenario in which devices contributing to the distributed learning process are not necessarily all present at the same time, but they may appear (and disappear) at different time instants. 
This configuration represents, for example, mobile social networking applications where users participate only when they are in a specific physical location (e.g., a gym, a park, a museum), and only for the time they spend there.  
 Therefore, in this case the learning phase in intended as a continuous process contributed by every device entering the location.
Note that in this case, in order to allow the continuous learning process, we need to assume that in the location there exist a permanent device (e.g. a totem) able to communicate with all the other devices, whose purpose is to store (and provide) to newly entered devices the information regarding the state of learning process executed so far. Or, it would be possible to use solutions like Floating Content~\cite{DBLP:journals/percom/OttHLKS11} in order to make the model available in the physical area by using only mobile nodes that move inside it.

In this dynamic scenario the learning process is done as follows. When a  number $s$ of devices enter into the location, they receive  the aggregate model $m$ that has been previously built by other devices (not necessarily still present) that executed the GTL learning procedure. The $s$ devices normally execute the GTL procedure, including in the distributed learning phase also the aggregate model $m$ . Once the learning phase is finished and a new  aggregate model $m'$ is obtained, the latter is combined with the previous version of the model  using the exponential moving average:
\begin{equation}
    m_{new} = \alpha m_{old} + (1-\alpha)m'
\end{equation}
where $0<\alpha\leq 1$ is a smoothing parameter used to balance the combination of the old model with the new one.  Note that, in this settings we assume that a single learning phase is atomic, i.e., during the learning phase devices involved in it do not change and no new devices can join.

We are interested in measuring i) the evolution of prediction performance of both GTL and noHTL  as long as devices arrive in the location to contribute to the continuous learning process and ii) the amount of traffic generated. Regarding the generated traffic, assuming that a permanent device, say $G$, exists that stores the models, we point out that in addition to the traffic related to each learning phase we have to include the traffic connected to sending the model $m$ to the $s$ devices at the beginning of the learning procedure, and the one for sending the $m'$ model to the permanent device . Formally, the additional traffic generated by the communication between $G$ and the $s$ devices is
\begin{equation}
    OH^{G} = d^{(0)}k(s+1)
\end{equation}
therefore in this scenario the total network overhead for GTL is 
\begin{eqnarray}
    OH^{dynGTL} = OH^{GTL}+OH^{G}
\end{eqnarray}
Conversely, for noHTL the way to compute the network overhead remains unchanged because the $s$ devices do not have to perform a retraining on their local data exploiting additional information contained in the aggregate model. In fact, their models are sent to the devices holding the aggregate model $m$ which computes the updated one ($m'$) and send it back to the $s$ devices. 

\begin{figure}
\centering
\subfloat[s=1]{
\includegraphics[width=0.5\textwidth]{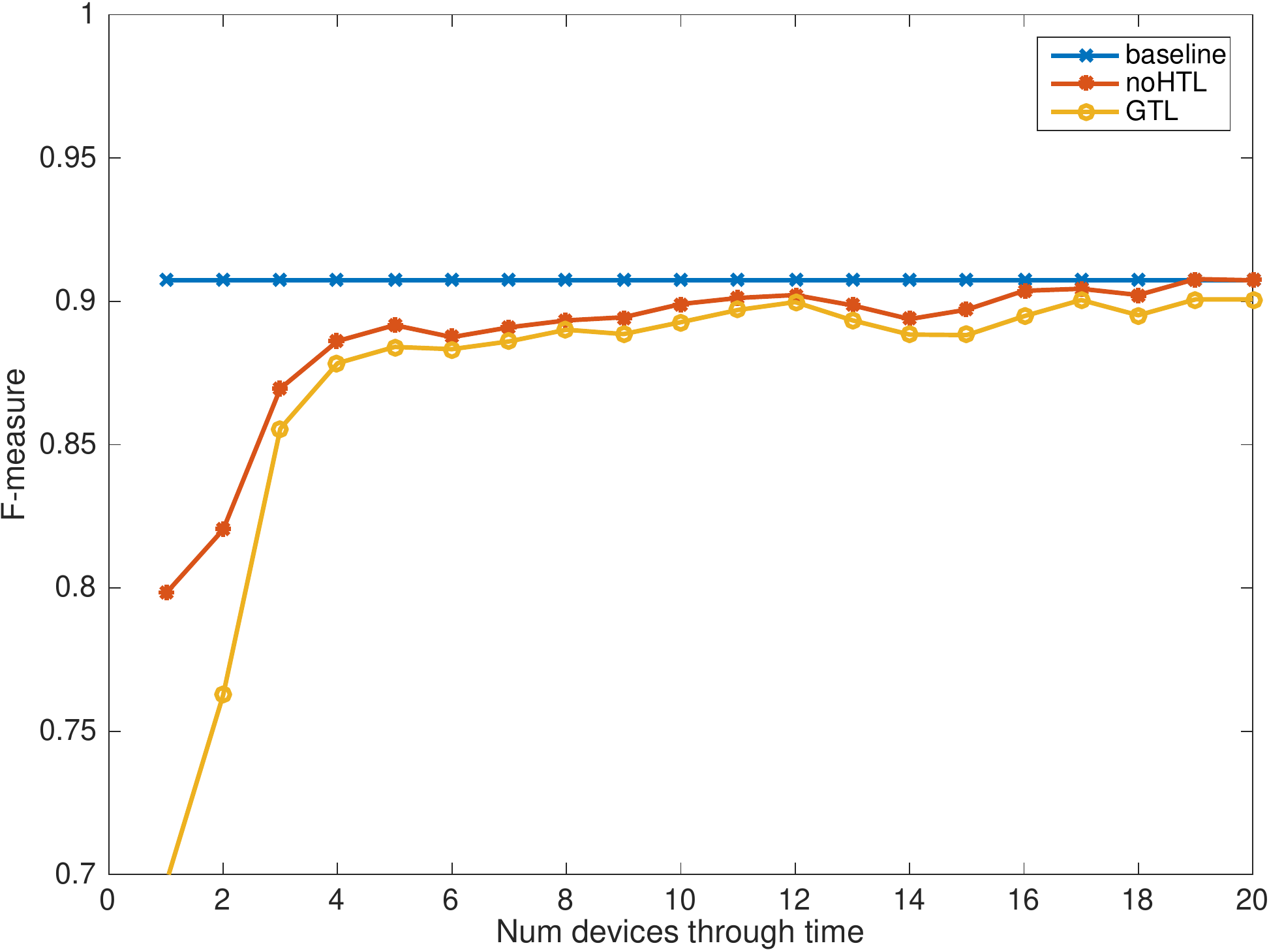}
}
\subfloat[s=4]{
\includegraphics[width=0.5\textwidth]{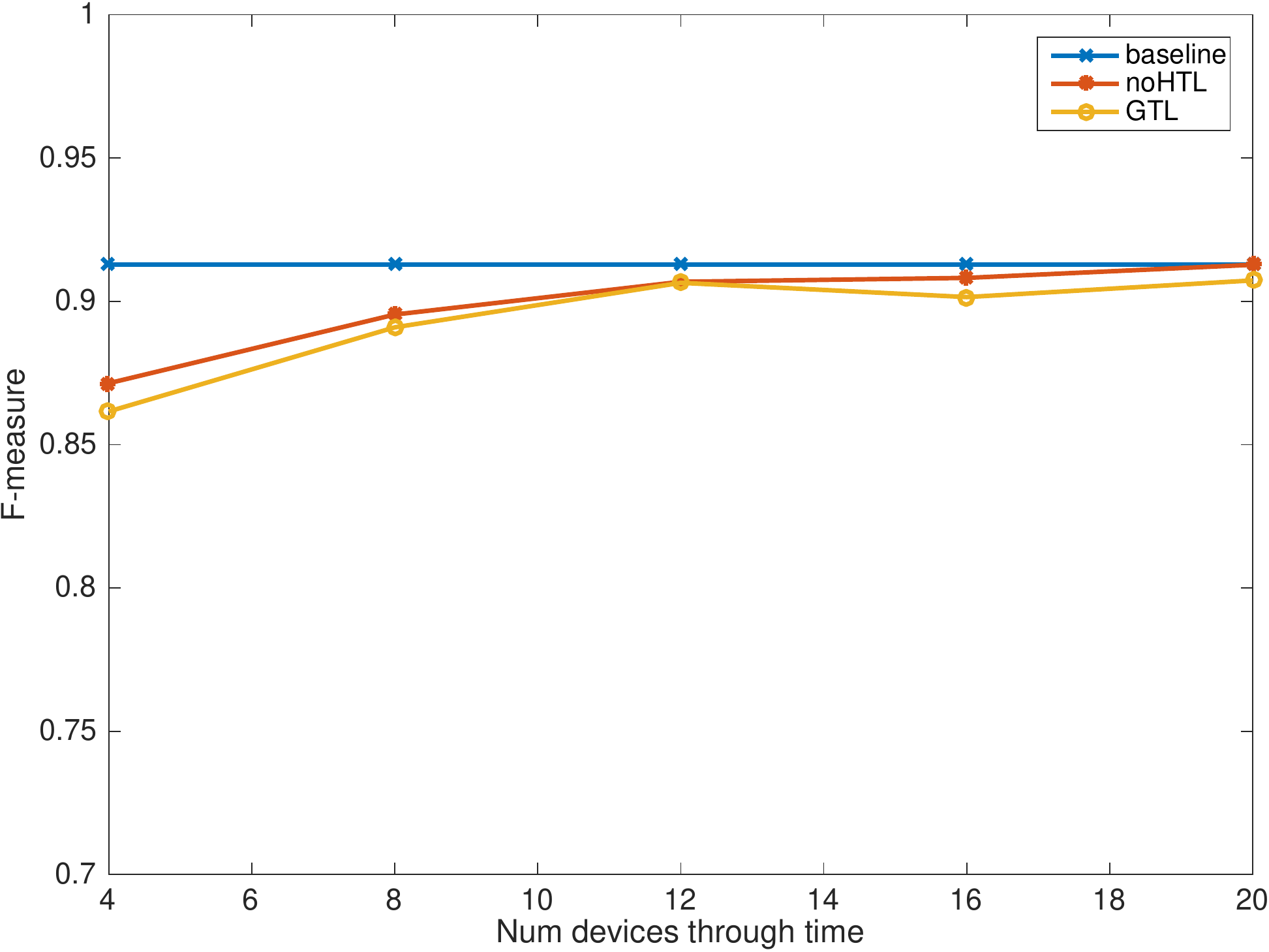}
}
\caption{Dynamic scenario. HAPT dataset. Prediction performance evolution with 1 and 4 new coming devices for each learning phase, respectively. \label{fig:dyn_hapt_acc}}
\end{figure}

\begin{figure}
\centering
\subfloat[s=1]{
\includegraphics[width=0.5\textwidth]{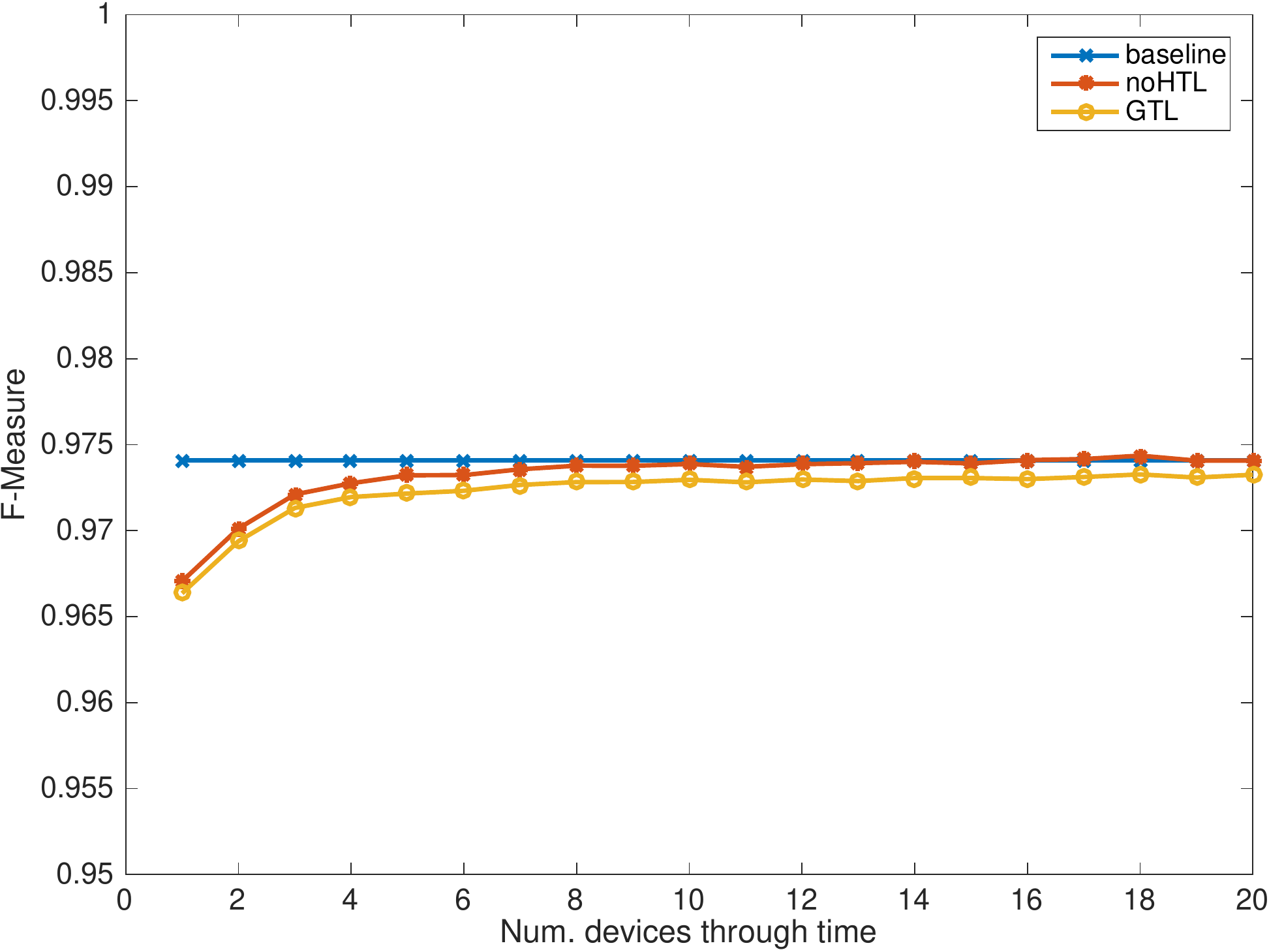}
}
\subfloat[s=4]{
\includegraphics[width=0.5\textwidth]{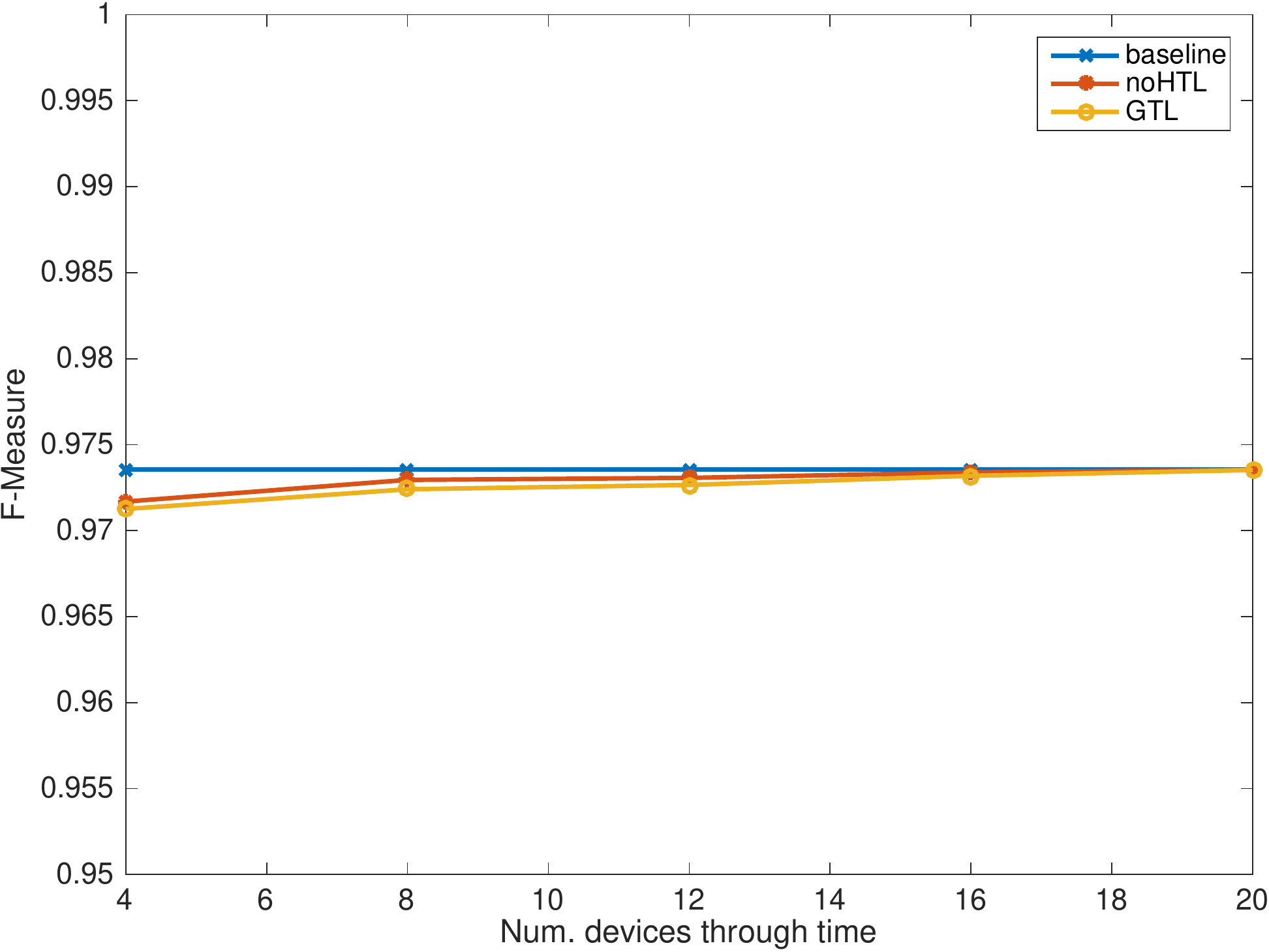}
}
\caption{Dynamic scenario. MNIST dataset. Prediction performance evolution with 1 and 4 new coming devices for each learning phase, respectively.\label{fig:dyn_mnist_acc} }
\end{figure}

Let us now present the performance of both GTL and noHTL in this dynamic scenario. Simulations are performed using both HAPT and MNIST datasets with balanced class distribution. In our simulations the number of devices entering the location at each new arrival is constant. For the sake of comparison we used as baseline the prediction performance obtained by noHTL in Sections \ref{sub:caseI} and \ref{sub:caseII}. In Figures \ref{fig:dyn_hapt_acc} and \ref{fig:dyn_mnist_acc} we show the prediction performance of both GTL and noHTL. As we can see, in both scenarios both approaches are able to converge after some time to the performance we would obtain running the same algorithms in presence of all device at the same time (see the baseline curve). Another interesting fact is that both GTL and noHTL, apart from the initial learning phases,  converge to the same performance. This suggests that in this specific case, using a learning mechanism based on Hypothesis Transfer Learning does not provide any significant advantage. We think that this could be due to the fact that the number of devices involved in each separate learning phase is so small that all the model they provide are useful to build the aggregate model and as a consequence none of them can be discarded.
\begin{table}
\centering
\caption{HAPT Dataset. Network overhead for different values of $s$.\label{tab:hapt_dyn_oh}}
\begin{tabular}{|c|c|c|c|c|c|}
    \hline
    & s &OH$^{noHTL}$ & OH$^{dynGTL}$ & Gain noHTL & Gain dynGTL \\
    \hline
    HAPT & 1 & $0.1$ MB &  $0.1$MB & $94 \%$ & $94\%$ \\
    HAPT& 4 & $0.4$ MB &  $1$MB & $94 \%$ & $84\%$ \\
    \hline
\end{tabular}
\end{table}

From the network overhead point of view, both approaches prove to be beneficial w.r.t. a cloud solution. Note that in this case the network overhead is computed averaging the traffic generated during each learning phase. The percentage gains are computed with respect to the amount of data held by devices involved in it (i.e., the amount of data that would be transferred in a Cloud-based approach).   
\begin{table}
\centering
\caption{MNIST Dataset. Network overhead for different values of $s$.\label{tab:mnist_dyn_oh}}
\begin{tabular}{|c|c|c|c|c|c|}
    \hline
    & s &OH$^{noHTL}$ & OH$^{dynGTL}$ & Gain noHTL & Gain dynGTL \\
    \hline
    MNIST & 1 & $0.05$ MB &  $0.05$MB & $98 \%$ & $98\%$ \\
    MNIST & 4 & $0.2$ MB &  $1$MB & $98 \%$ & $95\%$ \\
    \hline
\end{tabular}
\end{table}
Looking at Tables \ref{tab:hapt_dyn_oh} and \ref{tab:mnist_dyn_oh}, which report the average network overhead computed after each learning phase, we see that for both datasets, GTL and noHTL save more than $95\%$ of traffic. This results are in line with the ones presented in Section \ref{sec:noh} and confirm the efficiency of distributed learning, also in dynamic scenarios. 

\section{Conclusions} \label{sec:conclusions}
In this paper we study the performance of distributed learning solutions
for Fog Computing environments.
Precisely, we are interested in comparing the ability of the
different distributed learning approaches to accomplish a general machine learning task (such as
pattern recognition) on a dataset that is spread over several and separated
physical locations, without moving the data from where it is generated.

We have defined two types of algorithms, GTL and noHTL, respectively.
  The first one is based on the
  Hypothesis Transfer Learning framework. Nodes compute partial models based on
  local data, and then exchange all models between each other. They re-compute
  another model that combines all the ones received, exchange again the new
  models and aggregate them by averaging the models coefficient. On the other
  hand, noHTL is a simpler scheme where nodes compute the local
  models, exchange them, and aggregate them either through a standard
  consensus-based mechanism (i.e., again computing the average coefficients of
the received models), or based on a majority voting scheme.

We have analysed in detail the performance of these two algorithms both
  in terms of prediction performance and in terms of network overhead. We have
  used two reference datasets in the machine learning literature, i.e. the HAPT
and MNIST datasets. In all cases we have compared the performance of the
distributed learning solutions with those of a cloud-based solution where all
data have to be sent to a central cloud platform. We have considered both
balanced cases, i.e. where all nodes ``see'' the same distribution of data
points and all classes of data are equally represented, as well as unbalanced
cases, where either classes are unbalanced, or nodes see very different
distributions of data points. Our
results  show that adopting such kind of distributed
learning approaches we obtain prediction
performance comparable with a cloud-based solution, while saving between 60\% and 90\% of the network
traffic.
Among the two considered distributed solutions, GTL in several cases
yields better prediction performance, at the cost of higher traffic overhead.
However, we have also defined and evaluated a GTL solution where traffic can be
made almost equivalent to noHTL, without impacting on prediction performance.

Summarising, we think the results presented in this paper are very
  encouraging. They show that it is possible to analyse large-scale datasets
  collected by a large number of mobile devices, without necessarily
  transferring all these data to some central cloud platform, but through local
  analysis and appropriate distributed learning schemes. This is important,
  because the proposed solutions can drastically cut network traffic thus
  contributing to avoid congestion on wireless access networks. In addition, our
  algorithms lend themselves to cases where data needs to stay where it is
  collected for privacy and ownership reasons, but knowledge can be extracted
form it nevertheless.



\section*{Acknowledgment}

This work is partly funded by the European Commission under the H2020
REPLICATE (691735) , SoBigData (654024), and AUTOWARE
(723909) projects.

\section*{References}

\bibliography{bibliography}

\end{document}